 \documentclass{article}

 \usepackage{arxiv}
 \usepackage[utf8]{inputenc} 
 \usepackage[T1]{fontenc}    
\usepackage{graphicx}
\usepackage{relsize}
\usepackage{xfrac}
\usepackage{color}
\usepackage{amsfonts,amsmath,amssymb,bm}
\usepackage[sort&compress,numbers]{natbib}
\usepackage{xcolor}
\usepackage{multirow}
\usepackage{threeparttable,subcaption}

\usepackage{algorithm,tikz}
\usepackage[]{algpseudocode}
\usepackage{pgfplots}
\usepackage[colorlinks=true,citecolor=red,linkcolor=black]{hyperref}

\usetikzlibrary{matrix,chains,positioning,decorations.pathreplacing,arrows}
\usetikzlibrary{positioning,calc}
\usetikzlibrary{patterns,decorations.pathmorphing,decorations.markings}
\usetikzlibrary{shapes,arrows,chains,scopes}

\definecolor{snowymint}{HTML}{E3F8D1}
\definecolor{wepeep}{HTML}{FAD2D2}
\definecolor{portafino}{HTML}{F5EE9D}
\definecolor{plum}{HTML}{DCACEF}
\definecolor{sail}{HTML}{A3CEEE}
\definecolor{highland}{HTML}{6D885A}

\tikzstyle{signal}=[arrows={-latex},draw=black,line width=1.5pt,rounded corners=4pt]

\tikzstyle{block}=[draw=black,line width=1.0pt]
\tikzstyle{cell}=[style=block,draw=highland,fill=snowymint,
 rounded corners]
\tikzstyle{celllayer}=[style=block,draw,fill=portafino,
 inner sep=1pt,outer sep=0,
 minimum width=28pt, minimum height=14pt]
\tikzstyle{pointwise}=[style=block,ellipse,fill=wepeep,
 inner sep=1pt,outer sep=0, minimum size=12pt]

\tikzstyle{netnode}=[circle, inner sep=0pt, text width=22pt, align=center, line width=1.0pt]

\tikzstyle{inputnode}=[netnode, fill=yellow!20,draw=black]
\tikzstyle{hiddennode}=[netnode, fill=blue!20,draw=black]
\tikzstyle{outputnode}=[netnode, fill=red!20,draw=black]

\def\layerwidth{90pt}
\def\layerheight{14pt}

\tikzstyle{layer}=[style=block, draw, fill=black!20!white,
 inner sep=1pt,outer sep=0, font=\footnotesize,
 text centered, 
 minimum width=\layerwidth, minimum height=\layerheight]

\tikzstyle{fc}=[style=layer, fill=blue!30!white]
\tikzstyle{conv}=[style=layer, fill=green!30!white]
\tikzstyle{activation}=[style=layer, fill=orange!30!white]
\tikzstyle{pool}=[style=layer, fill=red!30!white]
\tikzstyle{bn}=[style=layer, fill=cyan!30!white]
\tikzstyle{recurrent}=[style=layer, fill=purple!30!white]
\tikzstyle{softmax}=[style=layer, fill=yellow!30!white]
\tikzstyle{point}=[]
\tikzstyle{branch}=[coordinate]

\def\vlayerwidth{30pt}
\def\vlayerheight{3pt}
\def\vblockheight{28pt}

\tikzstyle{vlayer}=[minimum width=\vlayerwidth, minimum height=\vlayerheight]
\tikzstyle{vblock}=[minimum width=\vlayerwidth, minimum height=\vblockheight, text width=1cm, align=center]

\colorlet{fn}{gray!90!green!30!white}
\colorlet{tp}{green!40!white}
\colorlet{fp}{red!40!white}
\colorlet{tn}{gray!90!red!20!white}

\usepackage{array}
\newcolumntype{P}[1]{>{\centering\arraybackslash}p{#1}}

\makeatletter
\renewcommand\paragraph{\@startsection{paragraph}{4}{\z@}%
 {3.25ex \@plus1ex \@minus.2ex}%
 {-1em}%
 {\normalfont\normalsize\bfseries}}
\makeatother

\newcommand{\Dtr}{\mathcal{D}_\mathrm{tr}}
\newcommand{\Dval}{\mathcal{D}_\mathrm{val}}

\newcommand{\Ntr}{{N}_\mathrm{tr}}

\newcommand{\Ii}{\mathcal{I}}
\newcommand{\Km}{\mathbf{K}}
\newcommand{\km}{\mathbf{k}}

\newcommand{\xm}{\mathbf{x}}

\newcommand{\ym}{\mathbf{y}}
\newcommand{\zm}{\mathbf{z}}

\newcommand{\xii}{\boldsymbol{\xi}}
\newcommand{\thetaa}{\boldsymbol{\theta}}
\newcommand{\sigmaa}{\boldsymbol{\sigma}}

\newcommand{\pinn}{\mathcal{M}_\mathrm{NN}}

\newcommand\Tstrut{\rule{0pt}{2.6ex}} 

\pgfplotsset{compat=1.16}
\begin{document}
\title{Prediction of ultrasonic guided wave propagation in solid-fluid and their interface under uncertainty using machine learning} 

 \author{
   Subhayan De\\
   Department of Aerospace Engineering Sciences\\
   University of Colorado Boulder\\
   Boulder, CO 80309 \\
   \texttt{Subhayan.De@colorado.edu} \\
   \And
  Bhuiyan Shameem Mahmood Ebna Hai\\
  Faculty of Mechanical Engineering\\
  Helmut Schmidt University\\
  Hamburg, Germany\\
  \texttt{ebnahaib@hsu-hh.de} \\
   \And
  Alireza Doostan\\
   Department of Aerospace Engineering Sciences\\
   University of Colorado Boulder\\
   Boulder, CO 80309 \\
   \texttt{doostan@colorado.edu} \\
    \AND
    Markus Bause \\
    Faculty of Mechanical Engineering\\
    Helmut Schmidt University\\
    Hamburg, Germany\\
    \texttt{bause@hsu-hh.de} \\
 }
%
%
\maketitle 
%
\begin{abstract}

Structural health monitoring (SHM) systems use non-destructive testing principle for damage identification. 
As part of SHM, the propagation of ultrasonic guided waves (UGWs) is tracked and analyzed for the changes in the associated wave pattern. 
These changes help identify the location of a structural damage, if any. We advance existing research by accounting for uncertainty in the material and geometric properties of a structure. 
The physics model used in this study comprises of a monolithically coupled system of acoustic and elastic wave equations, known as the wave propagation in fluid-solid and their interface (WpFSI) problem. 
As the UGWs propagate in the solid, fluid, and their interface, the wave signal displacement measurements are contrasted against the benchmark pattern. For the numerical solution, we develop an efficient algorithm that successfully addresses the inherent complexity of solving the multiphysics problem under uncertainty. We present a procedure that uses Gaussian process regression and convolutional neural network for predicting the UGW propagation in a solid-fluid and their interface under uncertainty. First, a set of training images for different realizations of the uncertain parameters of the inclusion inside the structure is generated using a monolithically-coupled system of acoustic and elastic wave equations. 
Next, Gaussian processes trained with these images are used for predicting the propagated wave with convolutional neural networks for further enhancement to produce high-quality images of the wave patterns for new realizations of the uncertainty. The results indicate that the proposed approach provides an accurate prediction for the WpFSI problem in the presence of uncertainty.

\end{abstract}
\keywords{Fluid-structure interaction \and Ultrasonic guided waves \and Gaussian process regression \and Super-resolution.}

\section{Introduction}

In the last few decades, maintenance and quality control of infrastructures have come to the forefront of engineering. As such, implementation of a damage identification system not only improves safety, but also minimizes the operational cost. Here, structural damages are defined as changes in the physical or geometric properties, boundary conditions, including those in the system connectivity, which adversely affect the performance of an infrastructure. In general, damage identification systems inform the user about any defect such as corrosion, delamination, or fatigue cracks in the structure in real time. This information can then be used for reactive measures. Additionally, these systems can also inform the user about the state of the structure, including an estimation of the remaining lifetime of the structure. 

Structural health monitoring (SHM) is the key component to implementing a damage identification system for a wide range of engineering infrastructures due to its ability to quickly respond to structural damages, improving life-cycle management and reliability by means of timely maintenance. The SHM system examines the state of structural health, providing the data for processing and interpretation, which can be used to estimate the remaining life-time of a structure. There are two types of SHM systems distinguished based on their operation: (a) active SHM, and (b) passive SHM \cite{SHMPWAS2014}. The active SHM system uses an approach similar to the one in non-destructive testing or evaluation (NDT or NDE) \cite{NDT2013,RAMSHM2017}, except that sensors in the active SHM system are permanently embedded into the structure \cite{SHMPWAS2014,CTSHM2011,SHMAS2016}. Based on the data provided by these sensors, the active SHM system may provide a recommendation for actions to be undertaken, e.g., to replace a specific part or repair the damage. 
While passive SHM does not require real-time data, the active SMH system is based on the continuous direct monitoring of the physical behaviour in real time \cite{CTSHM2011,UGW2011}. For a deeper discussion on SHM systems, we refer to \cite{NTSHM2013,NSTSHM2011,NTVSHM2010}, and the references therein. 

Currently, guided waves (GW) propagation approach is considered to be one of the most efficient NDT techniques. These techniques are sensitive to damage/inclusion, and can scan large areas. Accordingly, some studies considered ultrasonic guided waves (UGWs) \cite{UGW2014,GW2017,UGW2009,UGW2010}, while others Lamb wave propagation \cite{SHMAS2016,LW2016,LAMB1917}.
There is an extensive literature that addresses issues related to the GW-based SHM systems from an experimental perspective \cite{WAVE2014,LEE2014,REVIEW2007,REVIEW2010,REVIEW2016}. Historically, a GW-based SHM system has been represented by simulating a theoretical wave propagation model (e.g., Helmholtz, Lam\'{e}-Navier equations). Such an SHM system can identify and localize damage/inclusion by comparing the signal received at receiver sensors and the output of the physical model \cite{REVIEW2007,REVIEW2010,REVIEW2016,LWSHM2018}. However, there is a disproportionately small amount of modeling studies for this problem. This is despite the inherent advantage of mathematical modeling in structural damage identification. More specifically, it would allow evaluating a vast amount of alternative scenarios to prevent and manage any potential structural damage in a timely manner. The scarcity of studies, in part, can be explained by the associated modeling complexity. Recently, a subset of the authors \cite{EbnaHaiDE2017,EbnaHaiTS2019,EbnaHaiWpFSI2019,EbnaHaieXFSI2019} proposed a monolithically coupled system of acoustic and elastic wave propagation problem with or without fluid-structural interaction (FSI) effect to understand quantitatively and phenomenologically UGWs propagation and influence of the geometrical and mechanical properties of the inclusion and media. 

In the current paper, we focus on an ultrasonic wave propagation problem for a rigid solid plate in the absence of vibrations. That is, we consider the wave propagation in fluid-solid and their interface (henceforth, the WpFSI problem) \cite{EbnaHaiDE2017,EbnaHaiWpFSI2019}. To address the WpFSI problem in a rigid/ideal fluid-solid domain, we need to consider the fluid-solid interface effect throughout the modeling process, where the acoustic and elastic wave propagation models are fully coupled. 
Furthermore, the WpFSI problem considered in this paper allows gaining deeper understanding of the behavior of the UGWs propagation in solid-fluid, solid- solid, and their interface with or without any damage or inclusion through the solution of the associated nonlinear multiphysics problems. In particular, UGWs propagation pattern changes at the solid-fluid or solid-solid interface based on the number, location, size and material properties of the inclusion. However, due to the variations in external factors, such as material properties of the medium and inclusion, damage location, humidity, temperature \cite{TEMP2008,TEMP2009}, vibration \cite{EbnaHaieXFSI2019}, random noise, etc., uncertainties in UGWs propagation persist. Because of the complex nature of UGWs propagation, these uncertainties are challenging and computationally expensive to incorporate in a physical wave propagation model.

In the age of data, predictions using data-driven methods have found immense popularity. Among these methods is Gaussian process regression 
built from data and used for interpolation 
\cite{williams2006gaussian}. This approach for building a data-driven model is also known as \textit{kriging} and has been applied to a variety of scientific problems, such as, geostatistics \cite{krige1951statistical}, machine learning \cite{williams2006gaussian,koch2015efficient}, design optimization \cite{emmerich2006single,wang2007review,forrester2008engineering}, and so on. On the other hand, neural networks used for approximating functions \cite{goodfellow2016deep} have found recent popularity within the scientific community \cite{baker2019workshop}. 
In \cite{raissi2017physics,raissi2018hidden} neural networks are trained by minimizing a loss function that adds a contribution from the error in satisfying the governing differential equations. 
This approach was used for the propagation of seismic waves in  \cite{sun2021physics}. Convolutional neural networks and recurrent neural networks were used to simulate the wave dynamics in \cite{lino2020simulating} and \cite{sorteberg2018approximating}, respectively. For seismic wave simulation, \cite{moseley2018fast} used convolutional neural networks. 
The procedure to generate a high-quality image with more details from a low-quality image is known as the \textit{super-resolution} \cite{yang2019deep}. Convolutional neural networks were shown to perform this procedure efficiently in  \cite{dong2014learning,dong2015image}. 
\cite{stengel2019physics} used 
generative adversarial networks (GANs) \cite{goodfellow2016deep} for \textit{super-resolution} to generate high-resolution meteorological data from low-resolution images. 
GANs were trained in \cite{zhu2017wave} to generate images of the seismic wave propagation. \cite{melville2018structural,khurjekar2019accounting,rautela2021ultrasonic} used neural networks for UGW based damage identification and localization.  For a detailed list of publications on machine learning assisted modeling for physical problems, we refer the readers to \cite{willard2020integrating}. 

In uncertainty quantification of physical systems, 
\cite{zhang2019quantifying} used the dropout strategy \cite{hinton2012improving}, which ignores some of the connections in the networks with a probability for model and parametric uncertainty. 
Recently, \cite{de2020transfer} and \cite{de2020uncertainty} used transfer learning techniques and convolutional neural networks for uncertainty quantification of physical systems. 
However, these studies do not address the prediction of UGW patterns when uncertainty is present in a WpFSI problem. 
There are few studies on applied fluid-structure interaction (FSI) problems that use machine learning techniques to predict the flow. Among them, \cite{Miyanawala-FSI-ML2018} developed a hybrid data-driven technique for unsteady FSI problems, relying on the deep learning framework for long-term prediction of unsteady flow fields in a freely vibrating bluff body subjected to a wake-body synchronization. \cite{Whisenant-FSI-ML2020} employed neural networks in the Galerkin-free reduced-order modeling of the FSI applications.

As the solution of the associated multiphysics model for UGW propagation in solid-fluid and their interface imposes large computational costs, it limits the use of structural health monitoring methods for practical systems. To alleviate this issue, herein, we propose a method that improves on existing modeling approach by using Gaussian process regression and convolutional neural network for predicting UGW propagation in a solid, fluid, and their interface, in the presence of uncertainty in material and geometric properties. First, a set of training images for a small number of realizations of the uncertain properties of the fluid inclusion inside a solid plate is generated using a monolithically-coupled system of acoustic and elastic wave equations \cite{EbnaHaiWpFSI2019}. 
For a new realization of the uncertainty, we use an interpolation scheme using Gaussian process regression to generate an approximate image of the propagated wave through the plate. The quality of the propagated wave pattern is further improved through \textit{super-resolution} using convolutional neural networks. 
When validated with a separate dataset of propagating UGW, the results indicate that the proposed approach provides an accurate prediction for the WpFSI problem. 

\section{Background}

In this section, we first briefly describe the multiphysics model of the coupled wave propagation problem in the FSI domain, i.e., the WpFSI problem. Next, we introduce on Gaussian process regression and convolutional neural networks that are used in this paper to predict high-resolution images of propagating wave patterns.

\subsection{Physics Model} \label{sec:math_model}

\begin{figure}[!htb]
\centering
\includegraphics[scale=0.25]{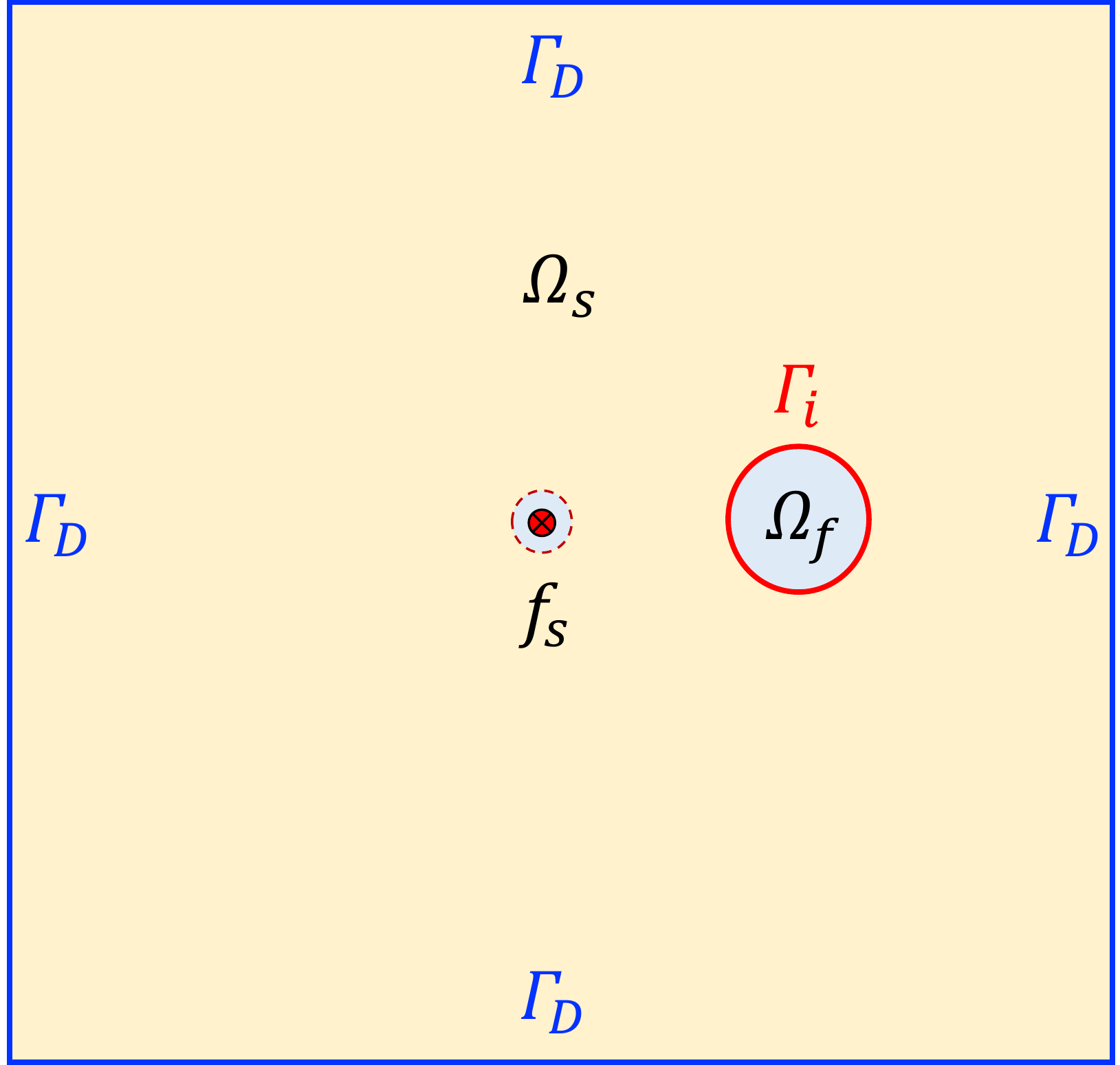}
\caption{Schematic representation of the WpFSI computational solid domain with a liquid inclusion at time $t=0$ (i.e., the reference domain). Here, $\Omega_s$ represents a solid plate with a fluid inclusion $\Omega_f$ with a common solid-fluid interface $\Gamma_i$. Homogeneous Dirichlet boundaries are defined by $\Gamma_D$ and $f_s$ represent a non-vanishing disc-shaped piezoelectric actuator.} 
\label{fig-wpfsi-domain}
\end{figure} 

Let $\widehat{\Omega}:=\widehat{\Omega}(t) \subset \mathbb{R}^d$, $d=\{2,3\}$ be a time-dependent bounded domain with Lipschitz boundary, $\widehat{\Gamma} \subset \mathbb{R}$. The computational domain $\widehat{\Omega}(t)$ is comprised of two time-dependent sub-domains $\widehat{\Omega}_{s}(t)$ and $\widehat{\Omega}_{f}(t)$, i.e. $\widehat{\Omega}(t)=\widehat{\Omega}_s(t) \cup \widehat{\Omega}_f(t)$, where, suffixes $``f"$ and $``s"$ are used to indicate the fluid and solid related terms, respectively. These two sub-domains do not overlap, i.e., $\widehat{\Omega}_{s}(t) \cap \widehat{\Omega}_{f}(t) = \emptyset$, and the corresponding solid-fluid common interface is denoted by $\widehat{\Gamma}_i= \widehat{\Gamma}_s \cap \widehat{\Gamma}_f \subset \mathbb{R}$. The outer boundaries $\widehat{\Gamma}_{D}$ are fixed (homogeneous Dirichlet boundary condition). Wave signal displacements and velocities on the outer boundaries are set to zero. For brevity, to highlight time-dependent variables, we use a circumflex symbol (i.e., `` $\widehat{}$ ''), and omit an explicit indication of the time-dependence in the rest of this paper.

The initial configuration (namely the reference domain at time $t=0$) is illustrated in Figure \ref{fig-wpfsi-domain}. In this paper, we denote the reference domain, boundaries, and common solid-fluid interface without any circumflex symbol. 
Furthermore, a non-vanishing disc-shaped piezoelectric actuator $f_s$ is embedded into the geometric center $(0,0)$ of a solid plate $ \Omega_s$. 
The corresponding outward normal vectors at the solid-fluid interface are henceforth denoted by $n_s$ and $n_f$. 
%
%
For the coupled elastic-acoustic wave propagation (WpFSI) problem in the solid or fluid domain \cite{EbnaHaiDE2017,EbnaHaiWpFSI2019}, the following spaces\footnote{
The standard Lebesgue space $L^p(X)$ where $1\leqslant p \leqslant \infty$ consists of measurable functions $u$, which are Lebesgue-integrable to the $p$-th power. The Sobolev space $W^{m,p}(X)$, $m \in N$, $1\leqslant p \leqslant \infty$ is the space of functions in $L^p(X)$ with distributional derivatives of an order up to $m$, and which belongs to $L^p(X)$. For $p=2$, $H^m(X) := W^{m,2}(X)$ is a Hilbert space equipped with the norm $\Vert \cdot \Vert_{H^m(X)}$. The Hilbert space with zero trace on $\Gamma_D$ is defined as $H_0^1(X)= \lbrace u \in H^1(X) : u\vert_{\Gamma_D} = 0\rbrace$.}
are employed as
		$L_{X}:=L^2(X)$, 
		$L_{X}^0:=L^2(X)/\mathbb{R}$, 
		$V_{X}:=H^1(X)$, 
		$V_{X}^0:=H^1_{0}(X)$, where $X$ can be used as $X := \Omega_s$ or $X := \Omega_f$. 
Since we use the variational-monolithic coupling method \cite{WickPFF2016,EbnaHaiDE2017,Richter2017book} to generate the training dataset, the wave signal displacements and velocity spaces are extended to the entire domain $\Omega = \Omega_s \cup \Omega_f$, which makes it more convenient to work with global $H^1$ function spaces. For the WpFSI problem, we work with the global $H^1$ function space
%
		$V_{ \Omega}^0:=\{ u \in H^1( \Omega)^d: u =0 \text{ on } \Gamma_D\}$.
%
Therefore, coupling conditions for the WpFSI problem in the variational formulation are automatically satisfied. Furthermore, for the time-dependent domains, we denote the corresponding spaces with a circumflex symbol (`` $\widehat{}$ '') notation. 

\begin{figure}[!htb]
	\begin{center}
		\includegraphics[scale=0.65]{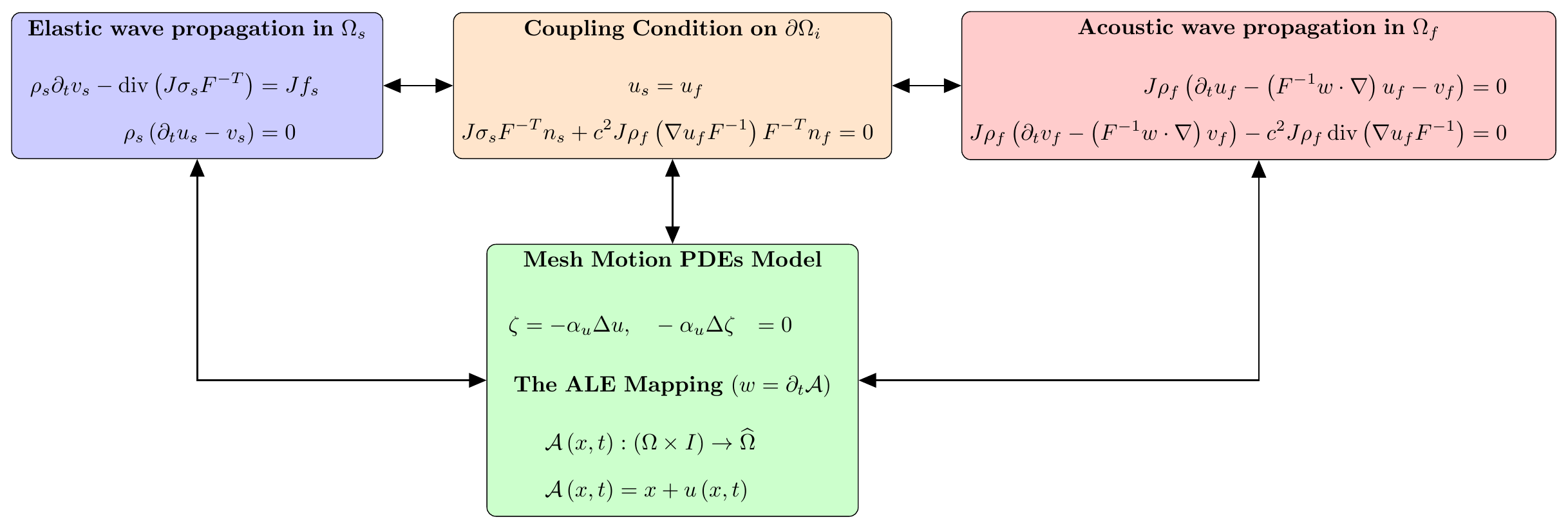}
		\caption{For the physics model of the WpFSI problem a monolithic coupling in the arbitrary Lagrangian-Eulerian (ALE) framework is used (see Appendix I). Here, $\rho$ is density, $u$ is the wave displacement, $v$ is the wave velocity, $c$ is the wave speed, $\zeta$ is auxiliary variables from mesh motion PDEs model (see Appendix II), ${F}$ is the deformation gradient, ${J}$ is the deformation determinant gradient, and suffixes $``s"$ and $``f"$ are used to indicate the solid and fluid related terms, respectively. } 
		\label{fig-wpfsi1}
	\end{center}
\end{figure}

Next, we introduce the vector-valued elastic wave equation for the UGWs propagation in the structure, and the acoustic wave propagation in terms of wave displacement in the fluid (see Appendix II). We formulate a symmetric system of equations, which is characterized by the displacement in both domains, for direct coupling of acoustic and elastic wave propagation problems (see Figure \ref{fig-wpfsi1}). 
In addition, to ensure that the problem is well-posed and is physically sensible we introduce coupling conditions. For the coupled formulation of the UGWs propagation problem in solid-fluid and their common interface, it is the balance between the normal components of displacements and forces on the common interface $\Gamma_i$ between the solid and fluid domains that needs to be ensured. This is achieved via the conditions
\begin{eqnarray}
\begin{aligned}
u_{s}&= u_{f} && \operatorname{on } \Gamma_i,\\
J \sigma_{s} F^{-T}n_s 
+ c^2 J \rho_f ( \nabla u_{f} F^{-1}) F^{-T}n_f
&=0 && \operatorname{on } \Gamma_i,
\end{aligned}
\label{eq4.79}
\end{eqnarray}
which relate to a Neumann-like boundary condition for the structure. To summarize, by combining the acoustics and elastic wave equations with the interface and boundary conditions, the coupled wave propagation problem in fluid, solid and their interface (cf. the WpFSI problem) is formulated. 
Here, the WpFSI problem describes the UGWs propagation in the computational domain using the principle of nondestructive evaluation. The associated variational form of the WpFSI problem is given below \cite{EbnaHaiWpFSI2019,EbnaHaiTS2019,EbnaHaieXFSI2019}.
%
\subsection{The Coupled Wave Propagation (WpFSI) Problem}
Find the global ultrasonic wave signal displacement $ u \in V_{ \Omega}^0$, velocity $ v \in L_{ \Omega}$, and auxiliary variables $ \zeta \in V_{ \Omega}$, such that the initial conditions $u(0)= u^0$ and $\partial_t u(0)= v^0$ are satisfied, and for almost all time $t \in I_t$ it holds that
\begin{eqnarray}
\begin{aligned}
( \rho \partial_t v, \phi^{v})_{ \Omega_s}
+( J \sigma_{s} F^{-T}, \nabla \phi^{v})_{ \Omega_s}
-( J f_{s}, \phi^{v})_{ \Omega_s}
+( J \rho \partial_t v, \phi^{v} )_{ \Omega_f}\\
-( J \rho ( F^{-1} w \cdot \nabla ) v , \phi^{v} )_{ \Omega_f}
+(c^2 J \rho ( \nabla u F^{-1}) F^{-T}, \nabla \phi^{v} )_{ \Omega_f}
&= 0 \quad \forall \phi^{v} \in V_{ \Omega}^0,\\
( \rho \partial_t u, \phi^{u})_{ \Omega_s}
-( \rho v, \phi^{u})_{ \Omega_s}
+ ( J \rho \partial_t u , \phi^{u})_{ \Omega_f}
- ( J \rho v, \phi^{u})_{ \Omega_f}\\
- ( J \rho ( F^{-1} w \cdot \nabla) u, \phi^{u})_{ \Omega_f}
+(\alpha_u \nabla \zeta, \nabla \phi^u)_{ \Omega_f}
- ( J f_a, \phi^u)_{ \Omega_f}
&=0 \quad \forall \phi^{u} \in L_{ \Omega},\\
(\alpha_u \zeta, \phi^\zeta)_{ \Omega_f}
- ( \alpha_u \nabla u, \nabla \phi^\zeta)_{ \Omega_f}
+ (\alpha_u \zeta, \phi^\zeta)_{ \Omega_s}
- ( \alpha_u \nabla u, \nabla \phi^\zeta)_{ \Omega_s}
&= 0 \quad \forall \phi^\zeta \in V_{ \Omega},
\end{aligned}
\label{eqwpfsi}
\end{eqnarray}
%
where $\phi^{u}$, $\phi^{v}$, and $\phi^\zeta$ are test-functions, and $w=\partial_t u_{f}$ is a mesh velocity driven by the external loads $f_a$. This allows us to link the WpFSI problem to the additional multiphysics problem, e.g. the FSI problem that can describe the solid deformation due to the ambient or internal fluid flow \cite{EbnaHaiWpFSI2019,EbnaHaiTS2019,EbnaHaieXFSI2019}. All the quantities can be found in Appendices I-III. 
The test-functions for the fluid and the solid sub-domains belong to the global test space $\phi^{v} \in V_{ \Omega}^0$, which implies that they coincide on the interface $\Gamma_i$. Accordingly, the Neumann coupling conditions at the interface are satisfied (in a variational way) and the following condition is implicitly accounted for in the WpFSI problem (see  \eqref{eqwpfsi}) as
\begin{eqnarray}
	\begin{aligned}
		\langle g_s, \phi^{v} \rangle_{ \Gamma_i} + \langle J g_{f} F^{-T}, \phi^{v} \rangle_{ \Gamma_i} &=0 && \operatorname{on } \Gamma_i.
	\end{aligned}
	\label{eq4.81}
\end{eqnarray}

\subsection{Gaussian Process Regression or Kriging} \label{sec:gp}

In Gaussian process regression, we assume that the measured data $\ym=[y_1,\dots,y_N]^T$ is generated according to a Gaussian process and can be given by
\begin{equation} 
\label{eqn:noisy_samples}
y_i = f(\xii_i) +\varepsilon_i,\qquad i = 1,\dots,N,
\end{equation}
where $\varepsilon$ is independent zero-mean Gaussian noise with variance $\sigma_n^2$. 
A zero-mean Gaussian process $f(\xii) \sim \mathcal{GP}\Big(0, \kappa(\xii,\xii^{\prime}) \Big)$ with covariance function $\kappa(\xii,\xii^{\prime})$ is written using a Gaussian distribution \cite{williams2006gaussian}
\begin{equation}\left\{
\begin{array}{c}
f(\xii) \\
f(\xii^\prime) \\
\end{array} \right\} \sim \mathcal{N}\left( \mathbf{0}, \left[ 
\begin{array}{cc}
\kappa(\xii,\xii) & \kappa(\xii,\xii^\prime) \\
\kappa(\xii^\prime,\xii) & \kappa(\xii^\prime,\xii^\prime)\\
\end{array}
\right]\right).
\end{equation}
Hence, we can write the joint probability density for prediction at $\xii^\prime$ as
\begin{equation}
\begin{split}
& \left\{
\begin{array}{c}
\ym \\
f(\xii^\prime) \\
\end{array} \right\} \sim \mathcal{N}\left( \mathbf{0}, \left[ 
\begin{array}{cc}
\Km+\sigma_n^2\mathbf{I} & \km_{f^\prime} \\
\km^{T}_{f^\prime} & \kappa_{f^\prime f^\prime}\\
\end{array}
\right]\right),
\end{split}
\end{equation}
where $\mathbf{I}$ is the identity matrix, $\Km$ is the $N\times N$ covariance matrix for the input $\{\xii_i\}_{i=1}^N$ with $\Km_{ij}=\kappa(\xii_i,\xii_j)$, $\km_{f^\prime}$ is the $N\times 1$ covariance vector between $\xii$ and $\xii^\prime$ with $\km_{f^\prime,i} = \kappa(\xii_i,\xii^\prime)$, and $\kappa_{f^\prime f^\prime} = \kappa(\xii^\prime,\xii^\prime)$. The posterior density of $f(\xii^\prime)$ is given by
$
f(\xii^\prime)\big\lvert \ym\sim \mathcal{N}\left( \bar{f}^\prime, \sigma^2_{f^\prime} \right),
$
%
where $\bar{f}^\prime=\km_{f^\prime}^{T}\left[ \Km+\sigma_n^2\mathbf{I} \right]^{-1}\ym$ and $\sigma_{f^\prime}^2 = \kappa_{f^\prime f^\prime} - \km_{f^\prime}^{T}\left[ \Km+\sigma_n^2\mathbf{I} \right]^{-1}\km_{f^\prime}$. 
\begin{center}
	\begin{threeparttable}[!htb]
		\caption{Examples of covariance kernels used in Gaussian process regression. } 
		\centering 
		\begin{tabular}{l l c c} 
			\hline 
			\Tstrut
			Kernel & $\quad\kappa(\xii, \xii^\prime)\quad$ & Hyperparameters \\ [0.5ex] 
			\hline 
			\Tstrut
			Radial basis & $\quad\gamma^2\exp\left(-\frac{\|\xii - \xii^\prime\|^2}{2\tau^2}\right)\quad$ & $[\gamma,\tau]$ \\ 
			Rational quadratic & $\quad\gamma^2\exp\left(1+\frac{\|\xii - \xii^\prime\|^2}{2\tau l^2}\right)^{-\tau}\quad$ & $[\gamma,l,\tau]$ \\
			Matern & $\quad \gamma^2\frac{2^{1-\nu}}{\Gamma(\nu)}\left(\frac{\sqrt{2\nu}\lVert\xii-\xii^\prime\rVert}{\tau}\right)^\nu B_{\nu}\left(\frac{\sqrt{2\nu}\lVert\xii-\xii^\prime\rVert}{\tau}\right)$ $^{*}\quad$ & $[\gamma,\tau,\nu]$ \\
			White noise & $\quad \gamma^2\delta_{\xii,\xii^\prime}$ $^{\dagger}\quad$ & $\gamma$ \\ [1ex] 
			\hline 
		\end{tabular}
		\label{tab:cov} 
		\begin{tablenotes}
			\item[$^*$] $\Gamma(\cdot)$ is the Gamma function; $B_\nu(\cdot)$ is the modified Bessel function of the second kind. 
			\item[$^\dagger$] $\delta_{\cdot,\cdot}$ is the Kronecker delta function.
		\end{tablenotes}
	\end{threeparttable}
\end{center}

There are many choices for the covariance function. Some of them that are tried in this paper are listed in Table \ref{tab:cov}. The hyperparameters in the kernel $\kappa(\cdot,\cdot)$ and the noise variance $\sigma_n^2$ are estimated from the measurement data by maximizing the likelihood function or its logarithm given by
\begin{equation}
\log{\mathrm{p}(\ym|\bm\theta_\mathrm{gp})} = -\frac{1}{2}\ym^T\left( \Km+\sigma_n^2\mathbf{I} \right)^{-1} \ym +\frac{1}{2}\log \lvert\Km+\sigma_n^2\mathbf{I}\rvert + \frac{N}{2}\log(2\pi),
\end{equation}
where $\thetaa_\mathrm{gp}$ is a vector consisting of the hyperparamters of the covariance kernel and $\sigma_n^2$.

\subsection{Neural Networks} \label{sec:nn}

Neural networks are popularly used for approximating functions in the scientific machine learning community. In its standard form (i.e., a feed-forward neural network), a neural network uses an affine transformation followed by a nonlinear one in every neuron. Multiple neurons are then arranged in layers which are then placed between the input and output layers. Hence, a neural network with $N_H$ hidden layers models the relation between the input $\xm$ and output $\ym$ as follows 
\begin{equation}
\ym = \mathbf{F}_0(\sigmaa(\mathbf{F}_{{N_H}}(\dots \sigmaa(\mathbf{F}_1(\xm;\thetaa_1)) \dots);\thetaa_{N_H});\thetaa_0), 
\end{equation} 
where $\{\thetaa_i\}_{i=1}^{N_H}$ are parameters of the $N_H$ hidden layers; $\thetaa_0$ is the parameter of the output layer; $\mathbf{F}_i$ for a standard neural network is an affine transform but for a convolutional neural network it involves a convolution operation and for a residual neural network involves inputs from previous layers; and $\sigmaa$ is a nonlinear function known as \textit{activation}. Most commonly used activation function is rectified linear unit (ReLU) given by $\max(0,z)$ for an input $z$. 
In this section, the convolution neural network used in this paper is briefly discussed next.

\subsubsection{Convolutional Neural Network (CNN)}
\begin{figure}
	\centering
	\includegraphics[scale=1]{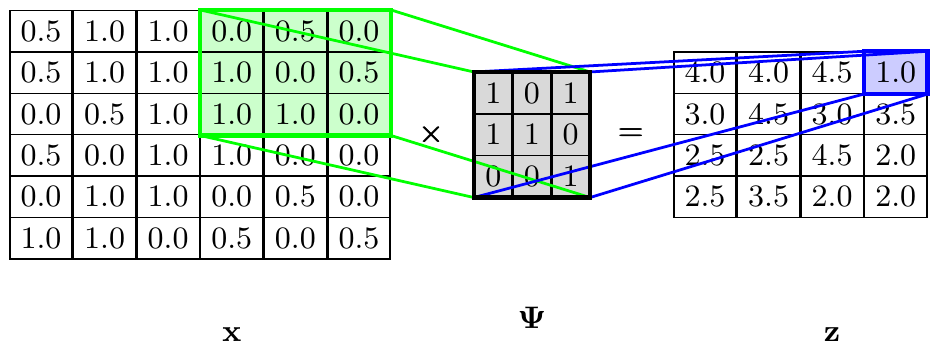}
	\caption{An example of cross-correlation as performed in (11).} 
	\label{fig:conv}
\end{figure}

\begin{figure}[!htb]
	\centering
	\includegraphics[scale=0.5]{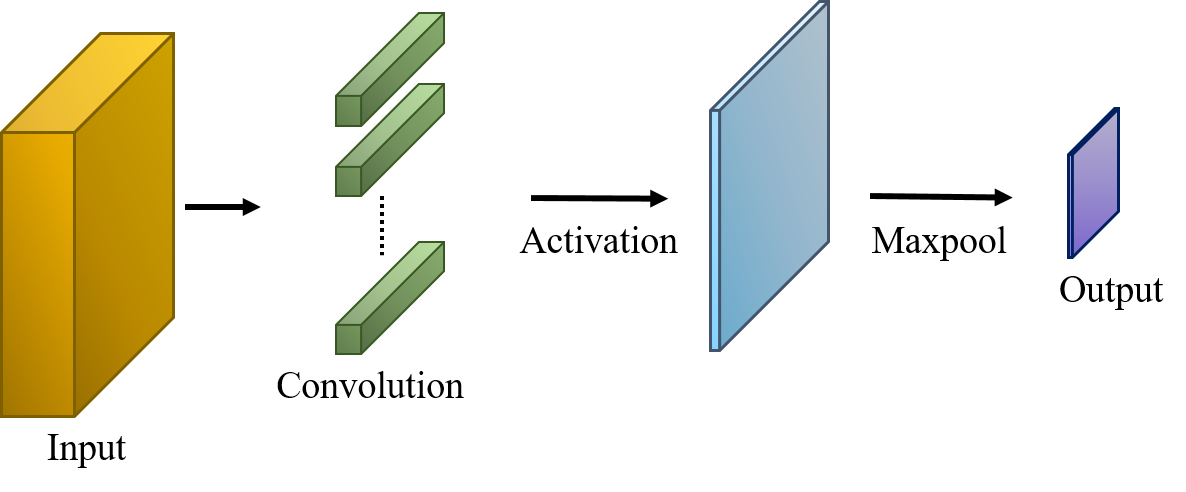}
	\caption{A typical single layer of convolutional neural network, where the operation in (11) is followed by the application of the activation function and a maxpooling operation. }
	\label{fig:cnn_1layer}
\end{figure}

In convolutional neural networks, a kernel $\Psi$ corresponding to a layer is learned for a two-dimenional input $\xm$ such that the output of the layer is given by
\begin{equation}\label{eq:conv}
z_{ij} = \sum_{q}\sum_r x_{i-q,j-r}\Psi_{qr}.
\end{equation} 
This representation reduces the number of parameters in the network for a kernel of size smaller than the size of the input resulting in sparse connectivity and reducing chances of overfitting.
Note that in software tools, such as PyTorch \cite{paszke2017automatic}, cross-correlation is used instead of the convolution, which is given by
\begin{equation}\label{eq:crosscor}
z_{ij} = \sum_{q}\sum_r x_{i+q,j+r}\Psi_{qr}.
\end{equation}
A schematic of this procedure is shown in Figure \ref{fig:conv}. 
Note that the kernel in \eqref{eq:crosscor} is two-times reflected version of the kernel in \eqref{eq:conv}. We denote the above operation in short using the $*$ symbol, \textit{i.e.}, $\zm=\xm*\boldsymbol\Psi$. 
Often the convolution in \eqref{eq:conv} or cross-correlation in \eqref{eq:crosscor} is followed by a \textit{maxpooling}, where the maximum over a window is selected to downsample the output $\zm$. In one layer of CNN, the convolution operation is followed by the use of an activation function and then the maxpool operation as shown in Figure \ref{fig:cnn_1layer}. 
In one variation of the network, known as the residual network or ResNet, the output from a previous layer is directly added to the output from the current layer. This architecture is shown to be effective for image processing \cite{tai2017image}. In this paper, we use this residual architecture of CNN to further enhance the images obtained from the Gaussian process regression. 


\subsubsection{Training of a Neural Network}
The mean-squared error between the prediction from the network and the training dataset $\Dtr = \{\xm_i,\ym_i\}_{i=1}^{\Ntr}$ given by
\begin{equation}
J=\frac{1}{\Ntr}\sum_{i=1}^{\Ntr} \Big\lVert\ym_i - \pinn(\xm_i;\thetaa) \Big\rVert^2,
\end{equation}
where $\pinn(\cdot;\cdot)$ is the prediction from the neural network,
is used as the cost function to train a neural network to optimize the parameters $\thetaa$. In the case of a CNN, the parameter consists of the kernel $\Psi$ and any bias if present. Stochastic gradient descent and its variants \cite{bottou2018optimization} are common choices for the optimization algorithm, where at $k$th iteration, we update the parameters as follows 
\begin{equation}
\thetaa^{(k+1)} \leftarrow \thetaa^{(k)} - \eta\frac{\partial J}{\partial \thetaa^{(k)}}.
\end{equation}
Here, $\eta$ is the step size also known as the learning rate. In this paper, we train the neural network used for \textit{super-resolution} using the Adam algorithm \cite{kingma2014adam,de2019topology}, a variant of the standard stochastic gradient descent. 


\subsection{Image Quality Assessment} \label{sec:im_quality}

In this paper, we compare the quality of the predicted images with the true results using the structural similarity (SSIM) index, a commonly used metric in image quality assessment. 
In \cite{wang2004image} the SSIM index between images $\Ii_1$ and $\Ii_2$ is defined as 
\begin{equation}
S(\Ii_1,\Ii_2) = \frac{(2\mu_1\mu_2+c_l)(2\sigma_{12}+c_c)}{(\mu_1^2+\mu_2^2+c_l)(\sigma_1^2+\sigma_2^2+c_c)}, 
\end{equation}
where $\mu_1$ and $\mu_2$ are the mean intensity of the images $\Ii_1$ and $\Ii_2$, respectively; $\sigma_1$ and $\sigma_2$ are the standard deviation of intensity of the images $\Ii_1$ and $\Ii_2$, respectively; and $c_l$ and $c_c$ are small constants to avoid division by zero.
This index has a range of $[-1,1]$ with $+1$ for a perfect match and $-1$ for an imperfect match. For images with multiple channels, an average over all the channels is used. 
We use this index to compare the images obtained from the proposed method with the true images in the validation dataset.

\section{Machine Learning based Framework for Prediction of Ultrasonic Guided Wave Propagation}\label{sec:framework}

In this paper, we propose a method for generating images of ultrasonic guided wave propagation at discrete time instances in the presence of an inclusion inside the medium. We further assume that there are uncertainties in the medium properties and the inclusion's geometry. The steps of the proposed machine learning based approach that includes the use of Gaussian process regression and deep neural networks are outlined next. 
Note that once trained, the proposed framework can produce high-resolution images of the propagated wave patterns for different realizations of the uncertain parameters at a negligible computational cost compared to solving the multiphysics model in the previous section. 

\subsection{Step I: Prediction of Field Variables using Gaussian Process Regression} 
A training dataset $\Dtr=\{\xii_i,\Ii_i\}_{i=1}^{\Ntr}$ of either images or field variables from simulations of the multiphysics model of the ultrasonic guided wave propagation is generated using the method described in the "Training and Validation Datasets" section, where each of these instances $\Ii$ has values at $H\times W$ coordinates and in total for $C$ field variables or channels. 
A family of Gaussian processes $\{\mathcal{GP}_{i}\}_{i=1}^{N_\mathrm{gp}}$ for $N_\mathrm{gp}=CHW$ is trained for predicting an $\Ii_\mathrm{pred}$ of the propagated wave for a realization of the uncertain variables $\xii_\mathrm{pred}$. 
Hence, the computational cost of this step is proportional to $CHW$. 
To avoid a large $CHW$ and significant prediction cost we only generate field variables in this step with moderate resolution. Then, in the next step, we use neural networks to generate more enhanced high-fidelity results in a procedure known as super-resolution as described below. 
Apart from reducing the computational cost of training the family of Gaussian process, super-resolution also helps in identifying the wave pattern better, as can be seen in the numerical examples. Further, it removes the requirement of generating high-quality images from a finer mesh for training the Gaussian processes. 
Note that a single GP for the whole domain does not produce accurate predictions for the WpFSI problem studied here and hence we choose to use a family of Gaussian process $\{\mathcal{GP}_i\}_{i=1}^{N_\mathrm{gp}}$. 

\subsection{Step II: Use of a Cascading Residual Network} \label{sec:stepII}

\begin{figure}[!htb]
	\centering
	\includegraphics[scale=0.45]{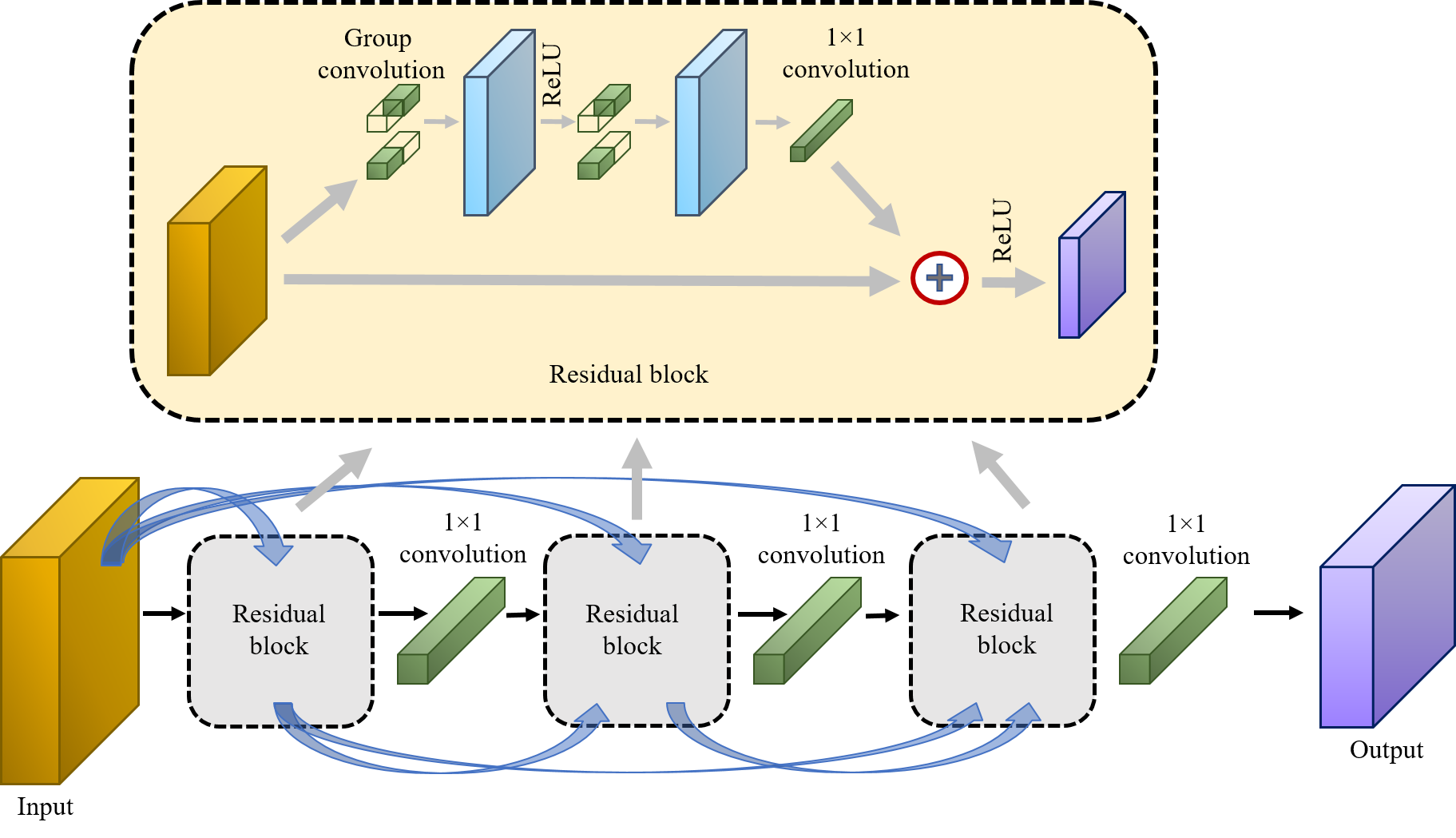}
	\caption{A schematic for the local cascading residual neural network used in Step II of the proposed method. Note that this cascade is used inside a global cascade (see \eqref{eq:global_cascade}). }
	\label{fig:CARN_schem}
\end{figure}

To further enhance the images and produce a high-fidelity prediction of the propagated wave pattern, the images obtained from Step I are further enhanced in this step using a procedure, known as the super-resolution (SR), with the help of a cascading residual network (CARN) following \cite{ahn2018fast}. Herein, a global cascading is used to generate the output image as 
\begin{equation} \label{eq:global_cascade}
\ym_{k} = \left[ \ym_0, \dots, \ym_{k-1}, \mathbf{F}^{\mathrm{local}}_{\!_{N_{l}}}(\ym_{k-1}) \right] * \Psi_{k} + b_{k}; \quad k=1,\dots,N_{g},
\end{equation} 
where $*$ denotes the convolution or cross-correlation operation as defined in the background section; $\ym_0= \xm*\Psi_0+b_0$ is the output from the first convolution operation; and $\mathbf{F}^{\mathrm{local}}_{\!_{N_l}}(\cdot)$ is the output from the local cascade. In the local cascade the output is 
\begin{equation}
\mathbf{F}_{j}^\mathrm{local} (\ym_{k}) = \left[ \mathbf{I}, \mathbf{F}_{0}^\mathrm{local}, \dots, \mathbf{F}_{j-1}^\mathrm{local}, \mathbf{F}^\mathrm{res}\left(\mathbf{F}_{j-1}^\mathrm{local}\right) \right] * \Psi_{j}^\mathrm{local} + b_j^\mathrm{local};\quad j=1,\dots,N_l,
\end{equation} 
where $\mathbf{F}_{0}^\mathrm{local}=\ym_k$; $\mathbf{F}^\mathrm{res}$ is the output from the residual layer; and $\Psi_{j}^\mathrm{local}$ and $b_j^\mathrm{local}$ are kernel and bias for the $j$th layer of the local cascade, respectively. 
Herein, the residual output is obtained after two convolution operations as follows 
\begin{equation}
\mathbf{F}_{}^{\mathrm{res}}\left( \mathbf{F}_{j}^\mathrm{local} \right) = \sigmaa\left(\dots\left(\sigmaa\left( \mathbf{F}_{j}^\mathrm{local}*\Psi_{1}^\mathrm{res} + {b}_{1}^\mathrm{res} \right)\dots\right)*\Psi_i^\mathrm{res}+b_i^\mathrm{res} + \mathbf{F}_{j}^\mathrm{local} \right);\quad i=1,\dots,N_r,
\end{equation} 
where $\Psi_j^\mathrm{res}$ and $b_j^\mathrm{res}$ are the kernel and bias of the $j$th convolution, respectively. Figure \ref{fig:CARN_schem} shows the local cascade with $N_l=3$ implemented herein that uses residual blocks with $N_r=3$ consisting of two $3\times 3$ group convolution followed by a $1\times1$ convolution as suggested in \cite{ahn2018fast}. The global cascade similarly use three $1\times1$ convolution (i.e., $N_g=3$) after every local cascade block. In the absence of a large quantity of high-quality images of the WpFSI problem, we use 1000 diverse 2K resolution (DIV2K) images from the dataset described in  \cite{Agustsson_2017_CVPR_Workshops}, a commonly used dataset to train super-resolution neural networks, to train the CARN. Note that the CARN needs to be trained once. The same network is used for all numerical examples in this paper. 

\subsection{Training and Validation Datasets} \label{sec:dataset}

To generate the training and validation datasets, we solve the WpFSI problem (i.e, \eqref{eqwpfsi} in the "Physics Model" section). 
In this work, we focus on an idealized (simplified) case, where the structure remains in its initial position and the acting force, e.g., the FSI effect, is negligible. Thereby, there is no solid mesh deformation due to the additional force. As a result, the deformation gradient and its determinant become $ F := I$, and $ J:=\det( F)=1$. The variational (or weak) formulation is subsequently prescribed in an arbitrary reference domain. In line with \cite{EbnaHaiDE2017}, \cite{Richter2017book}, and \cite{EbnaHaieXFSI2019}, we apply the Rothe method for the problem, where finite difference method is used for the temporal discretization and spatial discretization is performed based on a standard Galerkin finite element approach. 
More specifically, the time discretization is done by using the well-known shifted Crank-Nicolson scheme \cite{Richter2017book,WickMMT2011} with the time step size, $\Delta t=10$ $\mu$s and the total time $T=3$ ms.
Space-discretization is done via global refinement iteration 
into $67,450$ quadratic-mesh cells. Here, each cell has $73$ local degrees of freedom (DOFs) \cite{EbnaHaiTS2019}. 
\begin{figure}[!htb]
	\centering
	\begin{subfigure}[t]{0.35\textwidth}
		\centering
		\includegraphics[scale=1]{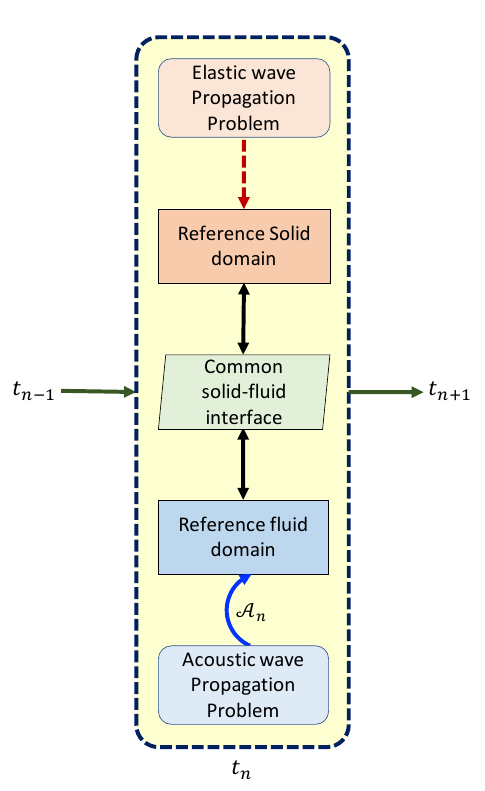}
		\caption{A monolithic coupling in the ALE framework at time $t_n$}
		\label{fig-alg1}
	\end{subfigure}
	\begin{subfigure}[t]{0.6\textwidth}
		\centering
		\includegraphics[scale=1]{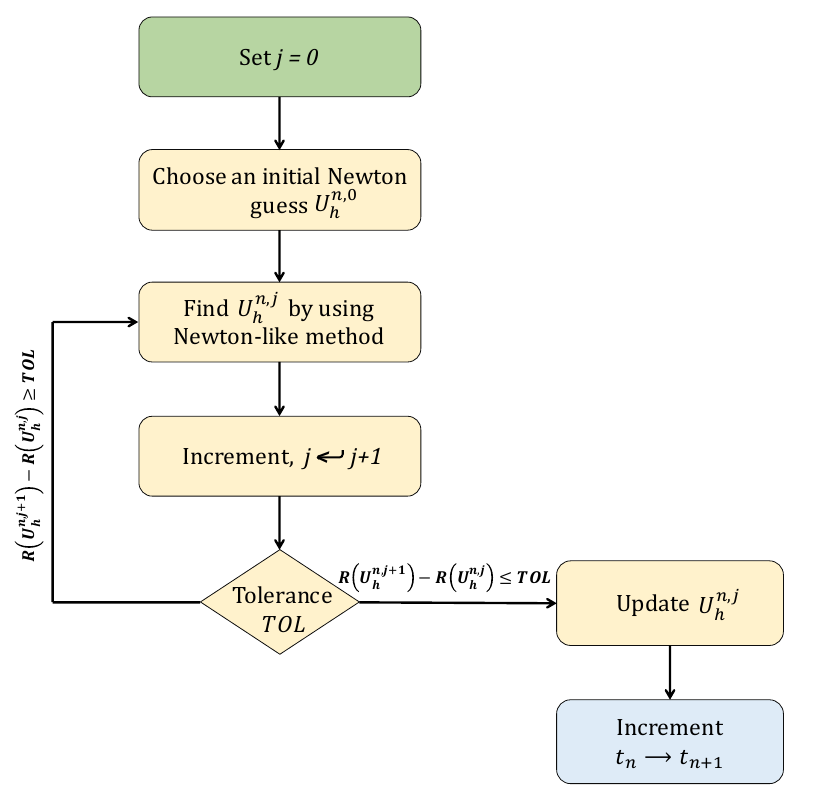}
		\caption{Solution technique with the Newton-like method}
		\label{fig-alg2}
	\end{subfigure}
	\caption{Schematic representation of a monolithic coupling and solution technique for the WpFSI problem. Here, in fully discretized form $ U_h^{n,j}=\{ u_h^{n,j}, v_h^{n,j}, \zeta_h^{n,j} \}$, and the nonlinear residual $R(\cdot)$ are defined in \cite{EbnaHaiDE2017,EbnaHaiWpFSI2019,EbnaHaieXFSI2019}.}
	\label{fig-alg}
\end{figure}
%
%
This nonlinear problem is solved by using a Newton-like method (see Figure \ref{fig-alg}). We 
apply the direct solver UMFPACK \cite{UMFPACK} for the solution of the linear system of equation at each Newton step (see Figure \ref{fig-alg}). 
The tolerance (TOL) for the residual-based Newton's method \cite{WickMMT2011,EbnaHaieXFSI2017} is set to $10^{-16}$. The implementation is accomplished by means of the software library package \textsc{DOpElib} \cite{DOpElib} and \texttt{deal.ii} \cite{deal}.

%
\begin{figure}[!htb]
	\centering
	\begin{subfigure}[t]{\textwidth}
		\centering
		\includegraphics[scale=0.78]{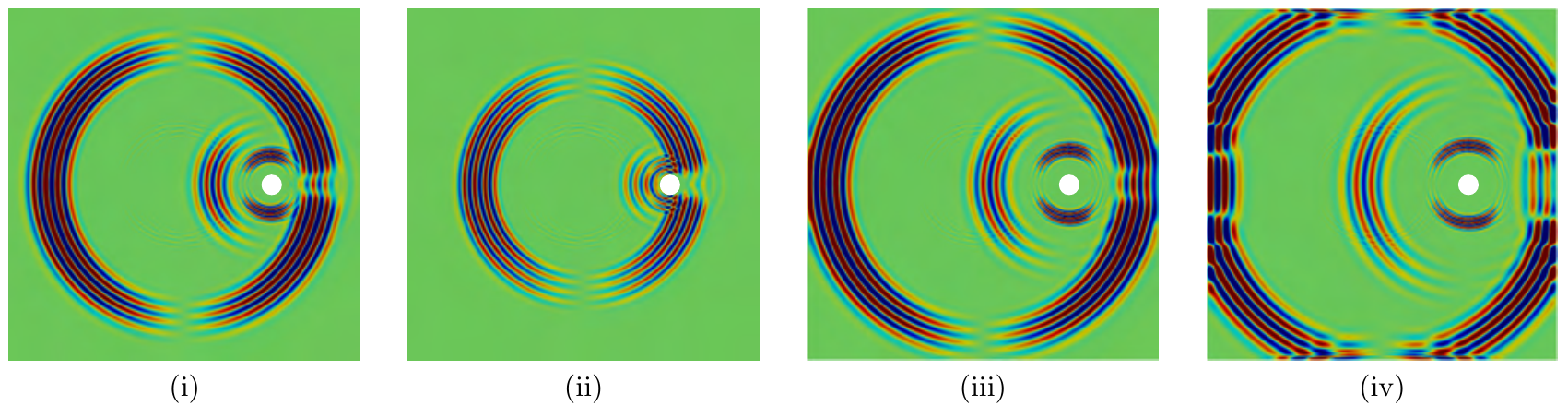}
		\caption{Wave propagation in a solid plate with a circular shape inclusion under uncertainties in the material properties (see Table \ref{tab:Ex_param}) at time, $t=3$ ms. (i) $E_{Y_s}=500$ kN/m$^2$, $\nu_s=0.4$, $\rho_s=1000$ kg/m$^3$ (ii) $E_{Y_s}=508.43$ kN/m$^2$, $\nu_s=0.3086$, $\rho_s=999.78$ kg/m$^3$ (iii) $E_{Y_s}=489.28$ kN/m$^2$, $\nu_s=0.3338$, $\rho_s=1002.48$ kg/m$^3$ (iv) $E_{Y_s}=510.33$ kN/m$^2$, $\nu_s=0.4489$, $\rho_s=1008.27$ kg/m$^3$ } \label{fig:7a}
	\end{subfigure}
	\\~\\~\\
	\begin{subfigure}[t]{\textwidth}
		\centering
		\includegraphics[scale=0.78]{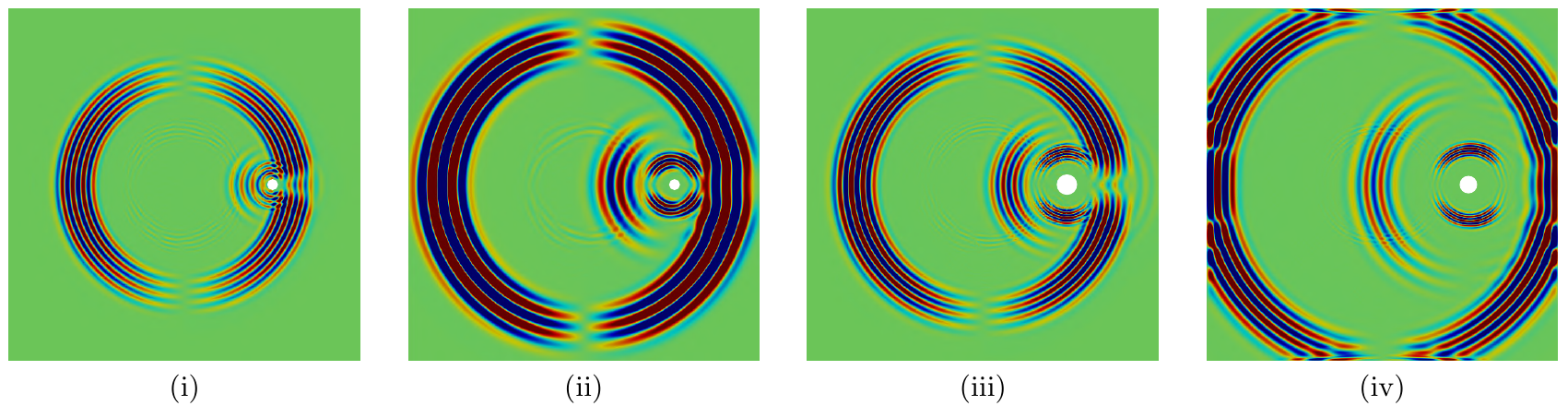}
		\caption{Wave propagation in a solid plate with a circular inclusion under uncertainties in the material properties (see Table \ref{tab:Ex_param}) and circular inclusion's radius, $r_c$ (see Table \ref{tab:Ex_dimension}) at time, $t=3$ ms. 
			(i) $r_c=0.0060$, $E_{Y_s}=489.28$ kN/m$^2$, $\nu_s=0.3338$, $\rho_s=1002.48$ kg/m$^3$ 
			(ii) $r_c=0.0053$, $E_{Y_s}=506.89$ kN/m$^2$, $\nu_s=0.4174$, $\rho_s=1001.53$ kg/m$^3$ 
			(iii) $r_c=0.0146$, $E_{Y_s}=501.19$ kN/m$^2$, $\nu_s=0.3878$, $\rho_s=1006.36$ kg/m$^3$
			(iv) $r_c=0.0085$, $E_{Y_s}=494.34$ kN/m$^2$, $\nu_s=0.4461$, $\rho_s=997.99$ kg/m$^3$} \label{fig:7b}
	\end{subfigure}\\~\\~\\
	\begin{subfigure}[t]{\textwidth}
		\centering
		\includegraphics[scale=0.78]{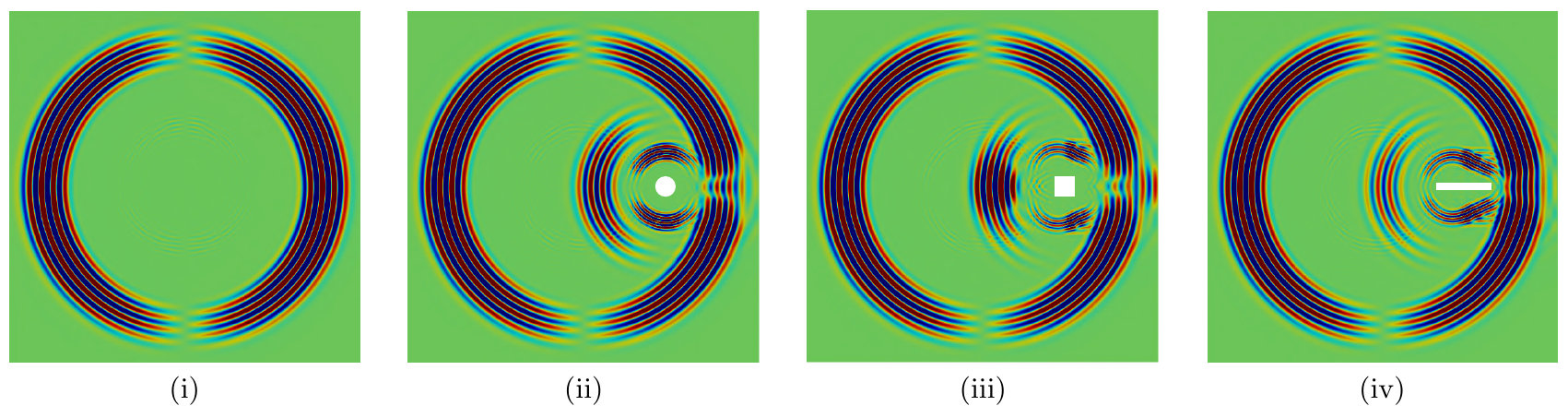}
		\caption{Wave propagation in a solid plate with uncertainties in the inclusion shape at time, $t=3$ ms. The material properties of the solid plate are $E_{Y_s}=500$ kN/m$^2$, $\nu_s=0.4$, $\rho_s=1000$ kg/m$^3$. (i) No inclusion, (ii) circular inclusion, (iii) square shape inclusion, and (iv) rectangular inclusion} \label{fig:7c}
	\end{subfigure}
	\caption{
	The wave displacement field component $u_x$ in a solid plate with a fluid inclusion.}
	\label{fig-wpfsi-results}
\end{figure}
%
%
The ultimate goal of tests conducted in this section is to present the tracking of the UGWs propagation in solid-fluid and their interface under uncertainty. The geometry of a square-shaped solid plate that is used for the numerical simulation is defined as follows. The plate has a length $L$ and a breadth $B$ aligned to the $x$-axis and the y-axis, respectively. Here, we assume $L = B = 0.35$ m. This plate is aligned in a Cartesian coordinate system; the solid domain is defined by $ \Omega_s:=(-0.175,-0.175) \times (0.175, 0.175)$. To foster comparison, the numerical simulations share the same configuration as above. Besides, all feature the non-vanishing disc-shaped piezo burst signal force $ f_s$ \cite{EbnaHaiWpFSI2019,EbnaHaieXFSI2017} (see Appendix III), 
which has a radius of $7.5$ mm and is located at the geometric center $(0,0)$ of the solid plate. For the purpose of visualisation of the wave displacement field component, a normalized color gradient with three channels: green, blue, and red is used. 
The displacements are normalized between $[-1,+1] $. For a schematic of different component of a propagated UGW patterns see Figure \ref{fig:signal1} in Appendix III.
%

Figure \ref{fig:7a} illustrates the wave signal displacement field component $ u_x$ of a five-cycle symmetric $S_0$ Lamb wave burst signal on the top surface (e.g., $xy$ plane) of a solid plate with a circular shape fluid inclusion to mimic mechanical damage under uncertainty in the base material properties. To understand the consequences of alternative base material properties, we compare results at a given time 
$t=3.0$ ms. 
In Figure \ref{fig:7a}, we depict the propagation of a symmetric wave package through a circular-shaped inclusion under four different base material specifications (e.g., $E_{Y_s}$, $\rho_s$, and $\nu_s$). Among these cases, we observe distinct wave propagation speeds, frequencies, wavelength, and amplitudes of the same Lamb wave burst signal as it passes from its origin to the boundary. 
In addition, as evident from Figure \ref{fig-wpfsi-results}, when wave signal approaches the fluid inclusion, it is partly reflected, part of it propagates further through the inclusion, turning into acoustic wave signals. It is worth noting that the wave signal reflections from the solid to a fluid exceed those from the fluid to the solid. Here, the magnitude of the reflected wave signal frequency is dependent upon the difference between the densities and the wave speeds of the materials; the wave signal reflections are relatively stronger for the stiffer structures. Due to these additional reflected wave signals implementation of machine learning technologies becomes more challenging. For more details about wave propagation behaviour in a solid-fluid domain, we refer the readers to \cite{EbnaHaiDE2017}, and \cite{EbnaHaiWpFSI2019}.
%
%

Next, Figure \ref{fig:7b} illustrates the wave propagation behaviour under uncertainty in the base material properties and the dimension of the circular-shaped fluid inclusion. In this case, the Lamb wave burst signal propagates with different wave propagation speed, frequency, amplitude, etc. In addition, we observe the reflection of an additional wave package with different patterns due to distinct dimensions of the inclusion. 
%
%
%
In Figure \ref{fig:7c}, we illustrated the displacement component $ u_{x}$ of a five cycle Lamb wave burst signal in a solid plate with different types of fluid inclusions. Here, the material properties of the fluid inclusion is characterized by density, $\rho_{f}=1\times10^3$ kgm$^{-3}$, and viscosity, $\nu_{f}=1\times10^{-3}$ m$^{2}$s$^{-1}$. Due to the different types of inclusions and different shapes of the solid-fluid common interface, the wave signal passes through the inclusion with distinct amplitudes and wavelengths. In addition, as the wave propagates from the base domain to the inclusion through the solid-fluid common interface, strong reflections occur. 

\section{Numerical Illustrations} 
Two numerical examples are used to illustrate the proposed approach in this paper. In the first example, we assume uncertainty is present in the material properties of the solid plate. In the second example, we further add uncertainty in the geometry of the inclusion. The dataset that we use here consists of images of the displacement component $u_x$ of the propagating wave. 

\subsection{Example I: Uncertainty in material properties} 
In our first example, the uncertainty is assumed in the material properties of the solid plate with known probability distribution functions as given in Table \ref{tab:Ex_param}. 
We use three different inclusion geometry, namely, circle, square, and rectangle to illustrate the proposed approach. The training dataset $\Dtr$ and the validation dataset $\Dval$, contain 50 and 30 sets of images for a time instance and a particular geometry of inclusion, which corresponds to (in total) 80 realizations of the material properties according to their probability distributions given in Table \ref{tab:Ex_param}. 
The training dataset $\Dtr$ is used to train the Gaussian processes $\{\mathcal{GP}\}_{i=1}^{N_\mathrm{gp}}$ for $N_\mathrm{gp}=CHW$. For these images, we have $C=3$, $H=W=270$ to keep the computational cost of Step I reasonable. 
On the other hand, the validation dataset $\Dval$ contains images that are used to compare the output of the proposed approach. We use a Matern kernel 
with $\gamma=1.0$, $\tau=1.0$, and $\nu=1.5$ (see Table \ref{tab:cov}) for the Gaussian process. In Step II, we use the neural network described in the previous section that is trained with 1000 open-source high-resolution images from the DIV2K dataset \cite{Agustsson_2017_CVPR_Workshops}. Note that the neural network is trained only once and is used for the next example as well. This trained network can further be used for future exercises. This step of performing a super-resolution on the Gaussian process output reduces the cost of Step I as well as remove the need for high-quality training images that are generated in this paper using the multiphysics model described in the background section to train the Gaussian processes. 

\setlength{\tabcolsep}{5pt}
\begin{center}
	\begin{threeparttable}[!htb]
		\caption{Specification of the uncertain parameters in Example I. } \label{tab:Ex_param} 
		\begin{tabular}{l l l l} 
			\hline 
			\Tstrut
			Parameter & Distribution & Mean & Std. Dev. \\ [0.5ex] 
			\hline 
			\Tstrut
			Elastic modulus, $E_{Y_s}$ & Truncated Gaussian$^*$ & 500 kN/m$^2$ & 10 kN/m$^2$ \\ 
			Poisson's ratio, $\nu_s$ & Uniform & 0.4 & 0.0577 \\
			Density, $\rho_s$ & Uniform & 1000 kg/m$^{3}$ & 5.77 kg/m$^{3}$ \\[1ex] 
			\hline 
		\end{tabular}
		\begin{tablenotes}
			\item[$^*$] Truncated below at zero. 
		\end{tablenotes}
	\end{threeparttable}
\end{center}

\begin{figure}[!htb]
	\centering
	\begin{subfigure}[t]{\textwidth}
		\centering
		\includegraphics[scale=1]{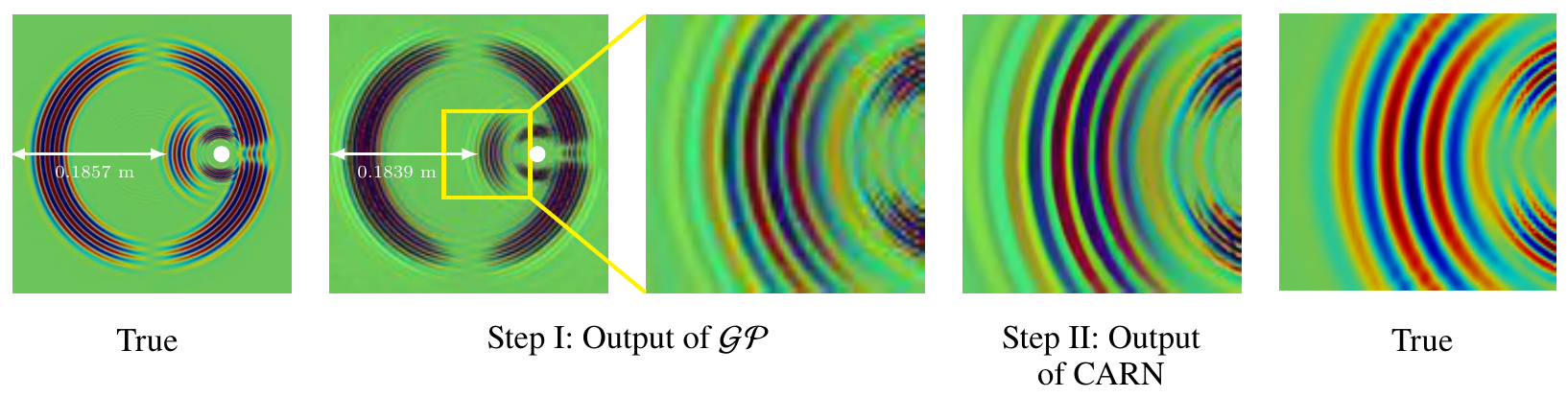}
		\caption{$E_{Y_s}=506.97$ kN/m$^2$, $\nu_s=0.3977$, $\rho_s=1000.5$ kg/m$^3$} \label{fig:circle_1}
	\end{subfigure} 
	\\~\\~\\
	\begin{subfigure}[t]{\textwidth}
		\centering
		\includegraphics[scale=1]{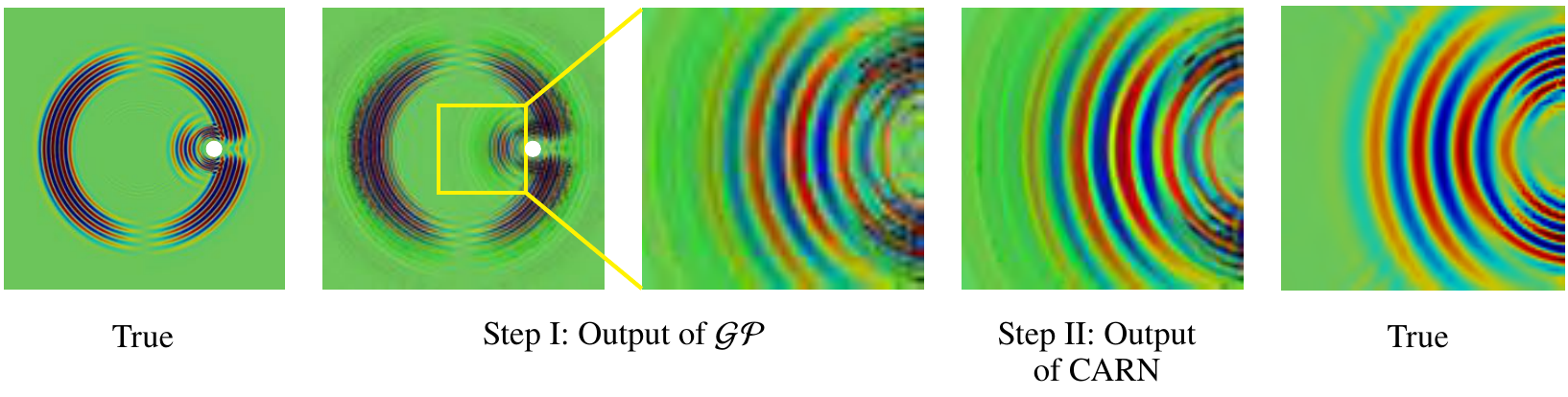}
		\caption{$E_{Y_s}=509.57$ kN/m$^2$, $\nu_s=0.3525$, $\rho_s=997.8$ kg/m$^3$}
	\end{subfigure}
	\caption{Prediction of the propagated UGW through a solid plate with uncertainties in the material properties at $t=3.0$ ms in the presence of a circular inclusion using the proposed approach for two realizations of the uncertainty.}
	\label{fig:circle}
\end{figure}

\begin{figure}[!htb]
	\centering
	\begin{subfigure}[t]{\textwidth}
		\centering
		\includegraphics[scale=1]{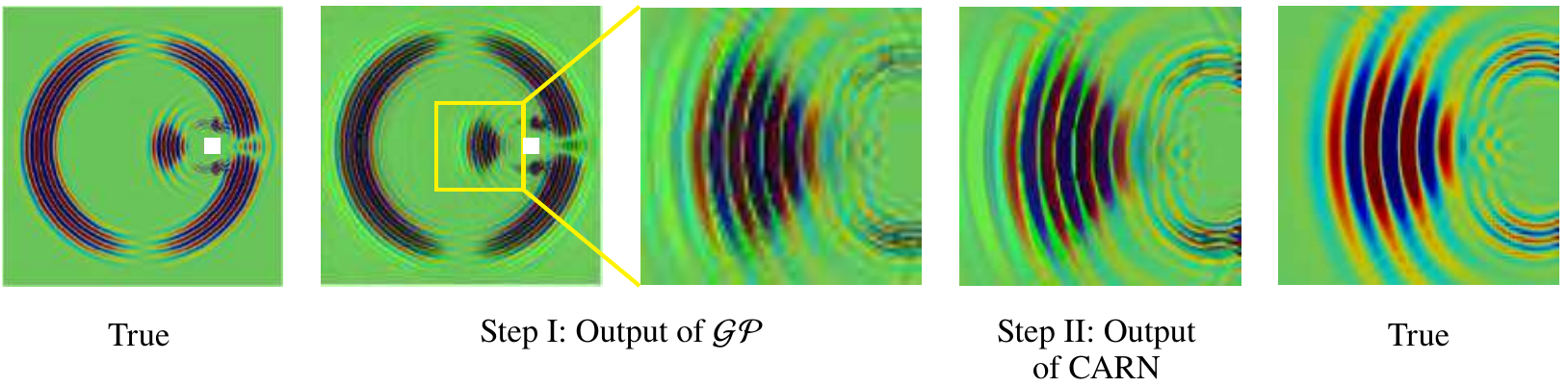}
		\caption{$E_{Y_s}=506.97$ kN/m$^2$, $\nu_s=0.3977$, $\rho_s=1000.5$ kg/m$^3$}
	\end{subfigure}
	\\~\\~\\
	\begin{subfigure}[t]{\textwidth}
		\includegraphics[scale=1]{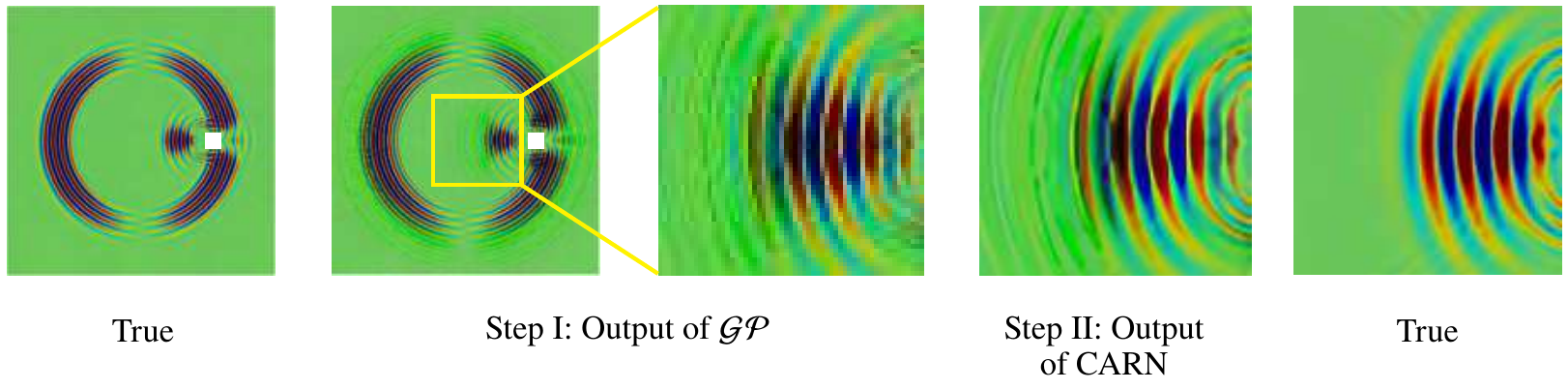}
		\caption{$E_{Y_s}=509.57$ kN/m$^2$, $\nu_s=0.3525$, $\rho_s=997.8$ kg/m$^3$}
	\end{subfigure}
	\caption{Prediction of the propagated UGW through a solid plate with uncertainties in the material properties at $t=3.0$ ms in the presence of a square shaped inclusion using the proposed approach for two realizations of the uncertainty.}
	\label{fig:square}
\end{figure}

\begin{figure}[!htb]
	\centering
	\begin{subfigure}[t]{\textwidth}
		\centering
		\includegraphics[scale=1]{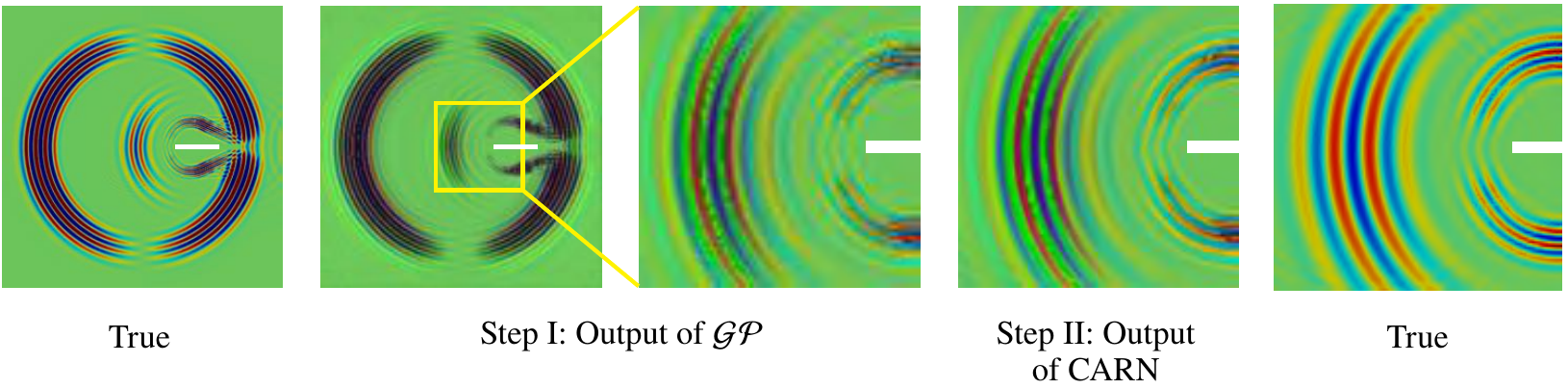}
		\caption{$E_{Y_s}=506.97$ kN/m$^2$, $\nu_s=0.3977$, $\rho_s=1000.5$ kg/m$^3$}
	\end{subfigure}
	\\~\\~\\
	\begin{subfigure}[t]{\textwidth}
		\centering
		\includegraphics[scale=1]{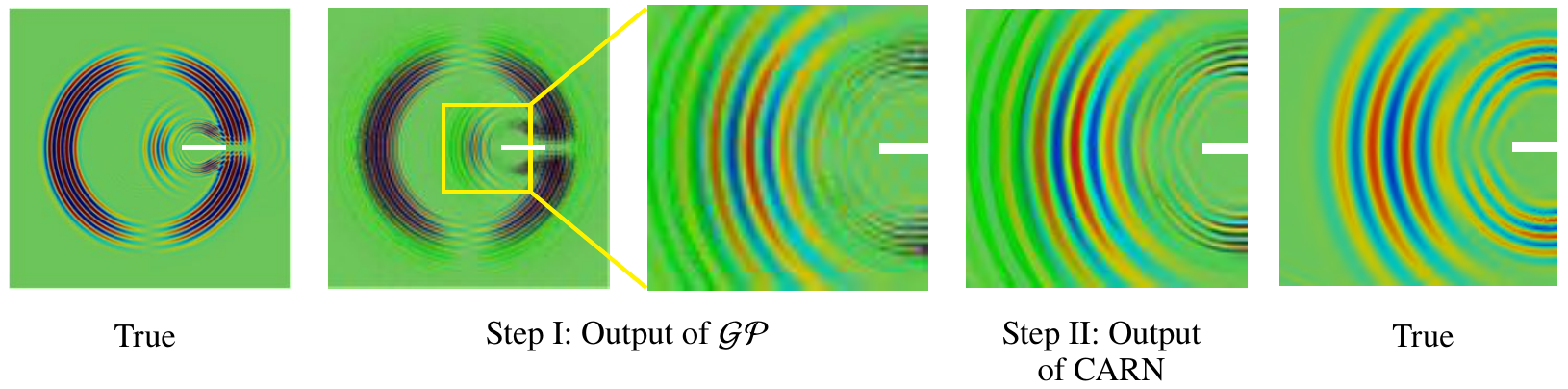} 
		\caption{$E_{Y_s}=509.57$ kN/m$^2$, $\nu_s=0.3525$, $\rho_s=997.8$ kg/m$^3$}
	\end{subfigure}
	\caption{Prediction of the propagated UGW through a solid plate with uncertainties in the material properties at $t=3.0$ ms in the presence of a rectangular inclusion using the proposed approach for two realizations of the uncertainty.}
	\label{fig:line}
\end{figure}

\begin{figure}[!htb]
	\centering
	\begin{subfigure}[t]{0.5\textwidth}
		\centering
		\includegraphics[scale=0.8]{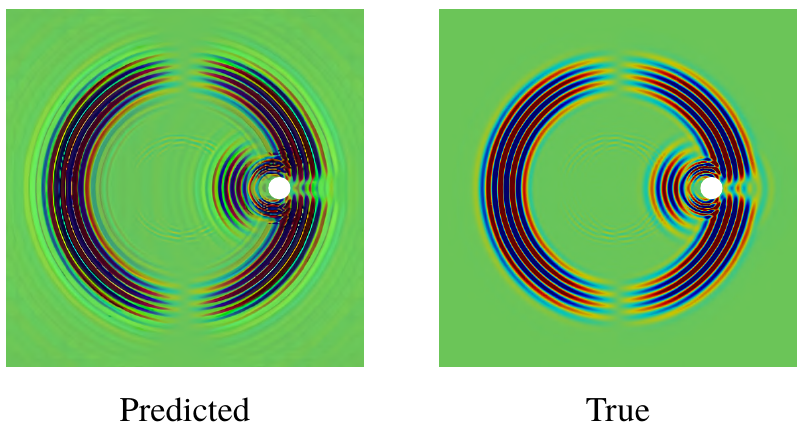}
		\caption{$t=2.7$ ms}
	\end{subfigure}\hfill
	\begin{subfigure}[t]{0.5\textwidth}
		\centering
		 
		\includegraphics[scale=0.8]{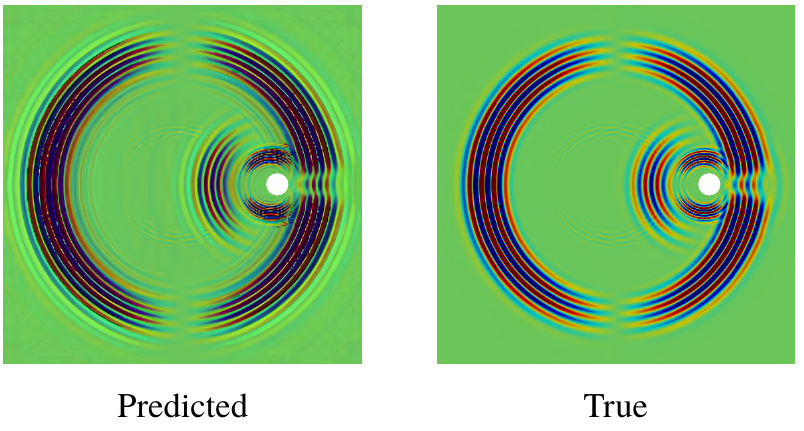}
		\caption{$t=3.0$ ms}
	\end{subfigure}
	\caption{Propagation of UGWs predicted using the proposed approach at two time instances for a realization of the uncertain parameters as $E_{Y_s}=505.10$ kN/m$^2$, $\nu_s=0.4017$, $\rho_s=1007.82$ kg/m$^3$.}
	\label{fig:time_lapse}
\end{figure}

\begin{table}[!htb]
	\caption{Image quality assessment using average SSIM (see the "Background" section) for the validation dataset $\Dval$. } \label{tab:ExI_im_quality} 
	\centering 
	\begin{tabular}{l c c c} 
		\hline 
		\Tstrut
		Inclusion shape~~ & ~~Average SSIM \\ [0.5ex] 
		\hline 
		\Tstrut
		Circle & 0.61\\ 
		Square & 0.61 \\
		Rectangle & 0.60 \\[1ex] 
		\hline 
	\end{tabular}
\end{table}

\subsubsection{Results:}

Figure \ref{fig:circle} compares the true images in the first and last columns for a circular inclusion with images from the Gaussian process regression (i.e., Step I) in the second column for two random realizations of the uncertain parameters. In the third column, we show an enlarged image of the Gaussian process output near inclusion to show the reflected wave. These images are of poor resolution due to the use of small $H$ and $W$. The fourth column shows the output from Step II, i.e., output from the CARN, which improves the image quality from Step I and shows the reflected wave in high resolution with two-fold increase in $H$ and $W$ each. 
Due to the uncertainty in the properties of the medium at $t=3.0$ ms the position of the wave including the reflected wave is different in these images. 
However, the proposed method shows high accuracy in the predicted images for the position of the wave as measured in Figure \ref{fig:circle_1}. 

Figures \ref{fig:square} and \ref{fig:line} show similar images for a square and a rectangular inclusion, respectively, for two realizations of the uncertainty. The propagated waves as predicted by the proposed method show a high level of similarity with the true images, as before. Table \ref{tab:ExI_im_quality} lists the average SSIM index for the validation dataset, which is around 0.6 for all three inclusions. Note that the SSIM index can vary from $-1$ for a mismatch to +1 for a perfect match. Hence, on average, the predicted wave patterns show reasonable agreement with their respective true images. To predict the propagation of wave for multiple time steps we repeat the procedure. Figure \ref{fig:time_lapse} shows the predicted images for $t=2.7$, and $3.0$ ms using the proposed method for one realization of the uncertain parameters and compares with the true wave patterns, which again shows reasonable agreement. The predictions for all these cases, however, show some errors near the zero displacement but the magnitude of these errors is small.  
Hence, this example shows that the proposed method can be used to efficiently predict the wave patterns for different shapes of inclusions and under uncertainty in medium's material properties and for multiple time-steps. 


\subsection{Example II: Uncertainty in material and geometric properties} 

In the second example, we increase the stochastic dimension of the problem by considering uncertainty in the inclusion geometry as well. In particular, we assume the diameter of the circular inclusion is uniformly distributed between 0.005 m and 0.015 m. The uncertainty in material properties of the solid plate are assumed same as before and listed in Table \ref{tab:Ex_dimension}. 
Two datasets, namely, the training dataset $\Dtr$ and the validation dataset $\Dval$, contain 50 and 30 sets of images for $t=3.0$ ms, for realizations of the geometry and material properties according to their probability distributions given in Table \ref{tab:Ex_dimension}. We used the training dataset $\Dtr$ to train the Gaussian processes $\{\mathcal{GP}\}_{i=1}^{N_\mathrm{gp}}$ for $N_\mathrm{gp}=CHW$. For these images, we have $C=3$, $H=W=270$ to keep the computational cost of Step I reasonable. A Matern kernel with $\gamma=1.0$, $\tau=1.0$, and $\nu=1.5$ is used for the Gaussian processes. 
We use these trained Gaussian processes to predict the propagation of waves and the same trained CARN from previous example for generating high-quality images. These predicted images are then compared with images in the validation dataset $\Dval$.

\begin{center}
	
	\begin{threeparttable}[!htb]
		\caption{Specification of the uncertain parameters in Example II. } \label{tab:Ex_dimension} 
		\begin{tabular}{l l l l l} 
			\hline 
			\Tstrut
			Part & Parameter & Distribution & Mean & Std. Dev. \\ [0.5ex] 
			\hline 
			\Tstrut
			Inclusion & Radius, $r_c$ & Uniform & 0.01 m & 0.0029 m \\ \hline \Tstrut
			\multirow{3}{*}{Solid plate} & Elastic modulus, $E_{Y_s}$ & Truncated Gaussian$^*$ & 500 kN/m$^2$ & 10 kN/m$^2$ \\ 
			& Possion's ratio, $\nu_s$ & Uniform & 0.4 & 0.0577 \\
			& Density, $\rho_s$ & Uniform & 1000 kg/m$^{3}$ & 5.77 kg/m$^{3}$ \\[1ex] 
			\hline 
		\end{tabular}
		\begin{tablenotes}
			\item[$^*$] Truncated below at zero. 
		\end{tablenotes}
	\end{threeparttable}
\end{center} 

\begin{figure}[!htb]
	\centering
	\begin{subfigure}[t]{\textwidth}
		\centering
		\includegraphics[scale=1]{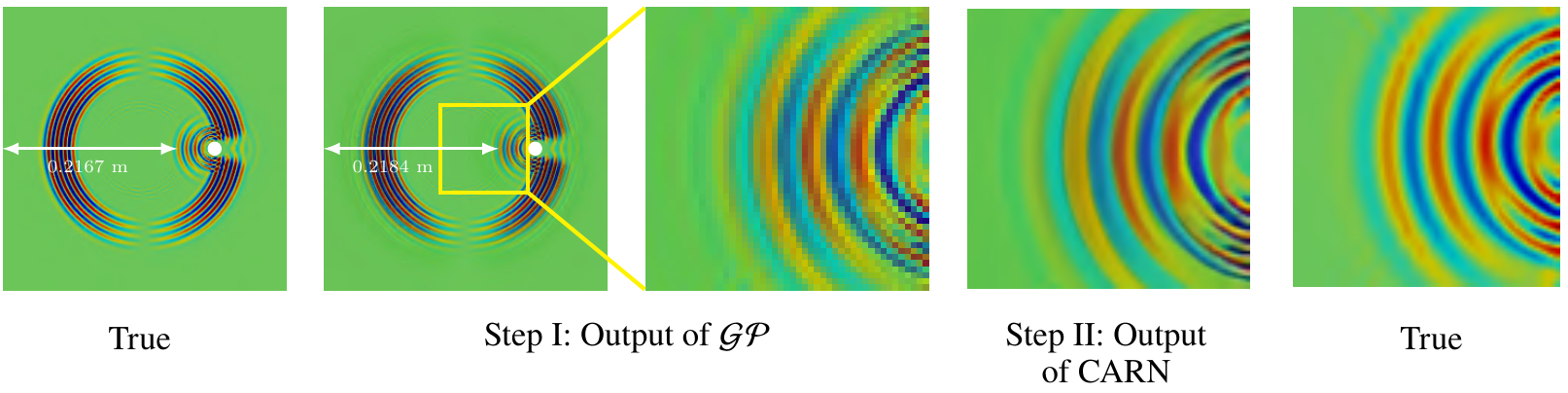}
		\caption{$r_c=0.0084$ m, $E_{Y_s}=489.13$ kN/m$^2$, $\nu_s=0.3222$, $\rho_s=1006.4$ kg/m$^3$} \label{fig:mat_size_a}
	\end{subfigure} \\~\\~\\
	  \begin{subfigure}[t]{\textwidth}
	  \centering
	\includegraphics[scale=1]{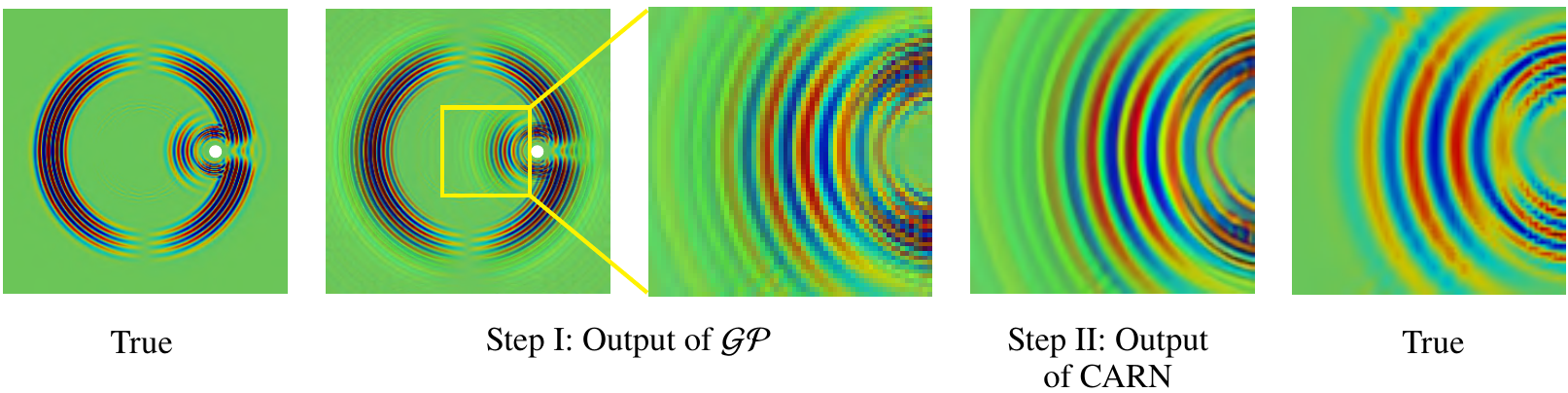} 
	\caption{$r_c=0.0075$ m, $E_{Y_s}=509.57$ kN/m$^2$, $\nu_s=0.3525$, $\rho_s=997.8$ kg/m$^3$} \label{fig:12b}
\end{subfigure}
\caption{Prediction of the propagated UGW through a solid plate with uncertainty in the material properties of the medium and radius of the circular inclusion at $t=3.0$ ms using the proposed approach for two realizations of the uncertainty.}
\label{fig:mat_size}
\end{figure}

\subsubsection{Results:} 
Figure \ref{fig:mat_size} compares the true images in the first and last columns with images from the Gaussian process regression (i.e., Step I) in the second column for two random realizations of the uncertain parameters. Since the computational cost of Step I is proportional to $CHW$ the resolution of the images obtained from Step I is low as illustrated in the third column with an enlarged image of the $\mathcal{GP}$ output near inclusion to show the reflected wave. The fourth column shows the output from Step II, i.e., output from the CARN, which improves the image resolution from Step I and shows the reflected wave better. In particular, the last two columns of Figure \ref{fig:12b} shows that the results from the proposed approach can even produce the wave patterns with more clarity than the solution of the multiphysics problem with a reasonable mesh resolution. 
Note that due to the uncertainty in the properties of the medium and inclusion at $t=3.0$ ms the position of the wave including the reflected wave is different for different realizations of the uncertain parameters in these images. However, the proposed method is able to predict the position of the reflected wave as shown in Figure \ref{fig:mat_size_a}. Again, there are some errors in the predicted images near zero displacements but the magnitude of this error is small, e.g., in Figure \ref{fig:12b}. For the validation dataset $\Dval$ the predicted images get an average SSIM value of 0.57, which is expected as we increase the stochastic dimension of the problem. 
Note that the SSIM index can vary from $-1$ for a mismatch to +1 for a perfect match. 
However, the proposed method shows high accuracy in the predicted images for the position of the wave. 
\section{Conclusions} 

In non-destructive tests, 
as part of structural health monitoring, the ultrasonic guided waves (UGWs) are tracked and 
contrasted against benchmark wave pattern, which is then used to inform about the presence of structural damages. However, if the properties of the medium are uncertain, the wave pattern can change for different realizations of the uncertainty. A multiphysics approach can be used to solve the associated mathematical model as discussed in the "Background" section but for many realizations of the uncertain parameters this requires significant computational budget. In this paper, we propose a two-step procedure that uses a Gaussian process regression first to predict the propagated wave pattern at a time instance. However, to keep the computational cost of this step reasonable the resolution of the predicted images are kept low. In the next step, we address this shortcoming by using a pre-trained neural network for increasing the resolution, a procedure known as super-resolution. Once trained, using this procedure the wave patterns can be estimated with a significantly small computational effort for many realizations of the uncertainty. We verified the accuracy of this procedure on two examples, where we assume uncertainty in the material and geometric properties of the medium and inclusions. Further, we 
considered three different shapes for the inclusions, namely, circle, square, and rectangle. The proposed method 
demonstrates high accuracy in predicting the wave patterns as well as the position of the reflected wave. In the future, the proposed method will be extended further with the use of a bi-fidelity dataset \cite{de2020bi}, to reduce the computational cost of training. 


\section{Data Availability Statement} 
Trained Gaussian processes and neural networks that support the findings of this study are available from the corresponding author upon reasonable
request. 

\section*{Acknowledgments}
This paper describes the technical results and analysis of the research conducted during the visit of the corresponding author to Smead Aerospace Engineering Sciences Department, University of Colorado, Boulder, CO, USA. The author would like to acknowledge his host Prof. Kurt Maute and support from the Helmut Schmidt University - University of the Federal Armed Forces, Hamburg, Germany. Any subjective views or opinions that might be expressed in the paper do not necessarily represent the views of the Federal Armed Forces of Germany. 

\nopagebreak
\nopagebreak
\section*{Appendix I. The Arbitrary Lagrangian-Eulerian (ALE) transformation}
\label{ALE}
The Arbitrary Lagrangian-Eulerian (ALE) mapping \cite{Richter2017book} is defined in terms of any mesh displacement $ u: \Omega \rightarrow \mathbb{R}^d$ such that 
$\mathcal{ A}( x,t): ( \Omega \times I) \rightarrow \widehat{\Omega}$, 
where $\mathcal{A}( x,t)= x + u(x,t)$. The ALE mapping is specified through the deformation gradient as 
$F:= \nabla \mathcal{ A}= I + \nabla u$, $J:=\det( F)$,
where $I$ is the identity matrix, and deformation determinant $ {J}$ is a function of $ {F}$. Moreover, function values in Eulerian and Lagrangian coordinates are given by
\begin{eqnarray}
\begin{aligned}
\widehat{x}= x(\widehat{x},t) + u(x, t) \qquad \Leftrightarrow \qquad \widehat{u}(\widehat{x}, t)= u( x, t) = \widehat{x}- x(\widehat{x},t), 
\end{aligned}
\label{eq6}
\end{eqnarray}
where $\widehat{x}=\mathcal{ A}( x,t)$. We define the ALE time derivative as follows \cite{Richter2017book}
\begin{eqnarray}
\begin{aligned}
\frac{\partial}{\partial t}\mathlarger{\Big\vert}_{\mathcal{ A}} \widehat{f} (\widehat{x},t)= \frac{\partial}{\partial t} \widehat{f} (\widehat{x},t)+ \widehat{w}\cdot\nabla \widehat{f}(\widehat{x},t),
\end{aligned}
\label{eq3.100}
\end{eqnarray}
where $ w( x,t)=\partial_t \mathcal{A}$ is the mesh/domain velocity for all $ x \in \Omega$ with $\widehat{w}(\cdot,t) = w(\cdot,t)\circ \mathcal{ A}^{-1}$. 
The mesh displacements are computed with the help of an additional partial differential equations (PDEs), which are referred to as mesh motion PDEs (or MMPDEs). We use the MMPDEs model based on the biharmonic equation. 
The alternative mesh motion techniques are elucidated in \cite{Richter2017book,WickMMT2011,Ciarlet1978book} and the references cited therein. 
The mixed formulation of the biharmonic mesh motion model with a control parameter $\alpha_u$ \cite{Ciarlet1978book} for the WpFSI problem is given by \cite{Richter2017book} 
%
\subsection{Biharmonic Mesh Motion Models}
Find an auxiliary variable $\zeta_f \in V_{ \Omega_f}$ and fluid displacement $ u_f \in V_{ \Omega_f, u_f}^0$, such that for almost all times $t \in I_t$ it holds that
\begin{eqnarray}
\begin{aligned}
(\alpha_u \nabla \zeta_f, \nabla \phi^u_f)
&=0 && \forall \phi^u_f \in V_{ \Omega_f, \phi^u_f}^0,\\
(\alpha_u \zeta_f, \phi^\zeta_f) - (\alpha_u \nabla u_f, \nabla \phi^\zeta_f)
&=0 && \forall \phi^\zeta_f \in V_{ \Omega_f},
\end{aligned}
\label{eq4.35}
\end{eqnarray}
%
where $\phi^u_f$ and $\phi^\zeta_f$ are test-function. This mixed formulation allows avoiding the use of $H^2$--conforming finite elements for the spatial discretization (see Ciarlet-Raviart mixed formulation \cite{Ciarlet1974BC}). We emphasize that the biharmonic model does not require a careful choice of a mesh-dependent parameter $\alpha_u$. Using this model, we simply choose a small number $\alpha_u > 0$ \cite{Richter2017book,WickMMT2011}. Let $\mathcal{ A}$ be a $C^1$-diffeomorphism; $\widehat{f} \in H^1(\widehat{\Omega}_f(t))$ be a differentiable function; and $\widehat{v} \in H^1(\widehat{\Omega}_f(t))^d$ a differentiable vector-field. Then the transformation between Eulerian and Lagrangian coordinate systems becomes \cite{Richter2017book}:
\begin{eqnarray}
\begin{aligned}
\partial_t \widehat{f} =\partial_t {f}-\left( {F}^{-1}\partial_t\mathcal{ A}\cdot {\nabla}\right) {f}, \qquad
d_t\widehat{f} =\partial_t {f}+\left( {F}^{-1}\left( v-\partial_t\mathcal{ A}\right)\cdot {\nabla}\right) {f},\\
\widehat{\nabla} \widehat{f} = F^{-T} \nabla f, \qquad \quad
\widehat{\nabla} \widehat{v} = \nabla v F^{-1}, \qquad \quad
(\widehat{v}\cdot \widehat{\nabla})\widehat{f}=\left( {F}^{-1} {v}\cdot {\nabla}\right) {f}.
\end{aligned}
\label{eq7}
\end{eqnarray}

\section*{Appendix II. Governing Equations}
\label{alleqs}

In this Section, we introduce the elastic wave equation for the UGWs propagation in the structure, and the acoustic wave propagation in terms of displacement in the fluid so that we can present a symmetric system of equations in the FSI domain; see  \eqref{eqwpfsi}. 
\subsection{Elastic Wave Equations in the Solid}
Find the linear elastic wave signal displacement $ u_{s} \in V_{ \Omega_s, \Gamma_i}^0$ and velocity $ v_{s} \in L_{ \Omega_s}$, such that the initial conditions $ u_{s} (0) = u_{s}^0$ and $\partial_t u_{s}(0)= v_{s}^0$ are satisfied, and for almost all time $t \in I_t$ it holds that
\begin{eqnarray}
\begin{aligned}
\rho_s(\partial_t u_{s} - v_{s})&=0 && \operatorname{in } \Omega_s,\\
\rho_s \partial_t v_{s}
- \nabla \cdot ( J \sigma_{s} F^{-T})
&= J f_{s} && \operatorname{in } \Omega_s,\\
u_{s} = u_{s}^D &=0 && \operatorname{on } \Gamma_{D_s},\\
( J \sigma_{s} F^{-T})n_s 
&= g_{s} && \operatorname{on } \Gamma_{N_s},
\end{aligned}
\label{eq-ewe1}
\end{eqnarray}
where $V_{ \Omega_s, \Gamma_i}^0:=\{ u_{s} \in H^1( \Omega_s)^d: u_{s} = u_{f} \text{ on } \Gamma_i, \hspace{0.3cm} u_{s} =0 \text{ on } \Gamma_D\}$, $ g_{s}$ is a vector-valued function, $ \rho_s$ is the density of the solid, and $I_t$ is the time interval. 
%
Accordingly, the variational (or weak) form of the elastic wave propagation problem in the Lagrangian coordinates is given by
%

\subsection{Elastic Wave Propagation in Lagrangian Coordinates}
Find the elastic wave signal displacement $ u_{s} \in V_{ \Omega_s, \Gamma_i}^0$ and velocity $ v_{s} \in L_{ \Omega_s}$, such that the initial conditions $ u_{s} (0) = u_{s}^0$ and $\partial_t u_{s}(0)= v_{s}^0$ are satisfied, and for almost all time $t \in I_t$ it holds that
\begin{eqnarray}
\begin{aligned}
( \rho_s (\partial_t u_{s} - v_{s}), \phi^{u}_s)_{ \Omega_s}
&=0&&\forall \phi^{u}_s \in L_{ \Omega_s},\\
( \rho_s \partial_t v_{s}, \phi^{v}_s )_{ \Omega_s}
+( J \sigma_{s} F^{-T}, \nabla \phi^{v}_s )_{ \Omega_s}
-\langle g_{s}, \phi^{v}_s \rangle_{ \Gamma_i}
-( J f_{s}, \phi^{v}_s)_{ \Omega_s}
&= 0&&\forall \phi^{v}_s \in V_{ \Omega_s}^0,
\end{aligned}
\label{eq4.70}
\end{eqnarray}
with the linearized stress tensor given by 
%
$\sigma_{s}= \frac{1}{ J} F \left( 2\mu_s E_{s}+\lambda_s\operatorname{ tr}( E_{s}) I\right) F^T$, 
$E_{s}=\frac{1}{2}\left( \nabla u_{s} F^{-1} + F^{-T} \nabla u_{s}^T\right)$,
%
where $\phi^{u}_s$ and $\phi^{v}_s$ are test-function; ${F}$ is the deformation gradient, and ${J}$ is the determinant of the deformation  gradient. Furthermore, $\mu_{s}$ and $\lambda_{s}$ are the Lam\'e coefficients for the solid. 
The relationship between the two material parameters $\mu_s$, $\lambda_s$, Poisson ratio $\nu_s$ and the Young modulus $E_{Y_s}$ reads: 
%
$\mu_s=\frac{E_{Y_s}}{2(1+\nu_s)}$, $\lambda_s=\frac{\nu_sE_{Y_s}}{(1+\nu_s)(1-2\nu_s)}$. 
%
%
%
\subsection{Acoustic Wave Propagation}
Find the acoustic wave signal displacement $\widehat{u}_{f} \in \widehat{V}_{\widehat{\Omega}_f}^0$ and velocity $\widehat{v}_{f} \in L_{\widehat{\Omega}_f}$, such that the initial conditions $\widehat{u}_{f} (0) = \widehat{u}_{f}^0$ and $\partial_t \widehat{u}_{f}(0)= \widehat{v}_{f}^0$ are satisfied, and for almost all time $t \in I_t$ it holds that 
\begin{eqnarray}
\begin{aligned}
\widehat{\rho}_f (\partial_t \widehat{u}_{f} - \widehat{v}_{f})
&=0 && \operatorname{in } \widehat{\Omega}_{f},\\
\widehat{\rho}_f \partial_t \widehat{v}_{f}
-c^2\widehat{\rho}_f \widehat{\nabla} \cdot (\widehat{\nabla} \widehat{u}_{f})
&= 0 && \operatorname{in } \widehat{\Omega}_{f},\\
\left(c^2\widehat{\rho}_f (\widehat{\nabla} \widehat{u}_{f})\right)\widehat{n}_f 
&= \widehat{g}_{f} && \operatorname{on } \widehat{\Gamma}_{N_f},
\end{aligned}
\label{eq-awq1}
\end{eqnarray}
where $g_{f}$ is a vector-valued function by the normal stress from the elastic wave propagation problem and $c$ is the wave speed.
%
Note, the elastic wave propagation problems \eqref{eq4.70} are formulated in the ALE framework (see Appendix I). Thereby, the acoustic wave equations \eqref{eq-awq1} have to be transferred to an arbitrary reference domain $ \Omega_f$. Accordingly, the variational formulation reads
%
\subsection{Acoustic Wave Propagation in the ALE Coordinates}
Find the acoustic wave signal displacement $ u_{f} \in V_{ \Omega_f}^0$ and velocity $ v_{f} \in L_{ \Omega_f}$, such that the initial conditions $ u_{f} (0) = u_{f}^0$ and $\partial_t u_{f}(0)= v_{f}^0$ are satisfied, and for almost all time $t \in I_t$ it holds that
\begin{eqnarray}
\begin{aligned}
( J \rho_f (\partial_t u_{f} - ( F^{-1} w \cdot \nabla) u_{f} - v_{f}), \phi^{u}_f)_{ \Omega_f}
&=0&&\forall \phi^{u}_f\in L_{\Omega_f},\\
( J \rho_f \partial_t v_{f}, \phi^{v}_f )_{ \Omega_f}
-( J \rho_f ( F^{-1} w \cdot \nabla ) v_{f} , \phi^{v}_f )_{ \Omega_f} &\\
+(c^2 J \rho_f ( \nabla u_{f} F^{-1}) F^{-T}, \nabla \phi^{v}_f )_{ \Omega_f}
-\langle J g_{f} F^{-T}, \phi^{v}_f \rangle_{ \Gamma_i}
&= 0&&\forall \phi^{v}_f\in V_{\Omega_f}^0.
\end{aligned}
\label{eq4.77}
\end{eqnarray}
%

\section*{Appendix III. The non-vanishing Lamb wave burst signal}
\label{a1}

\begin{figure}[!htb]
	\centering
	\begin{subfigure}[t]{\textwidth}
		\centering
		\includegraphics[scale=0.9]{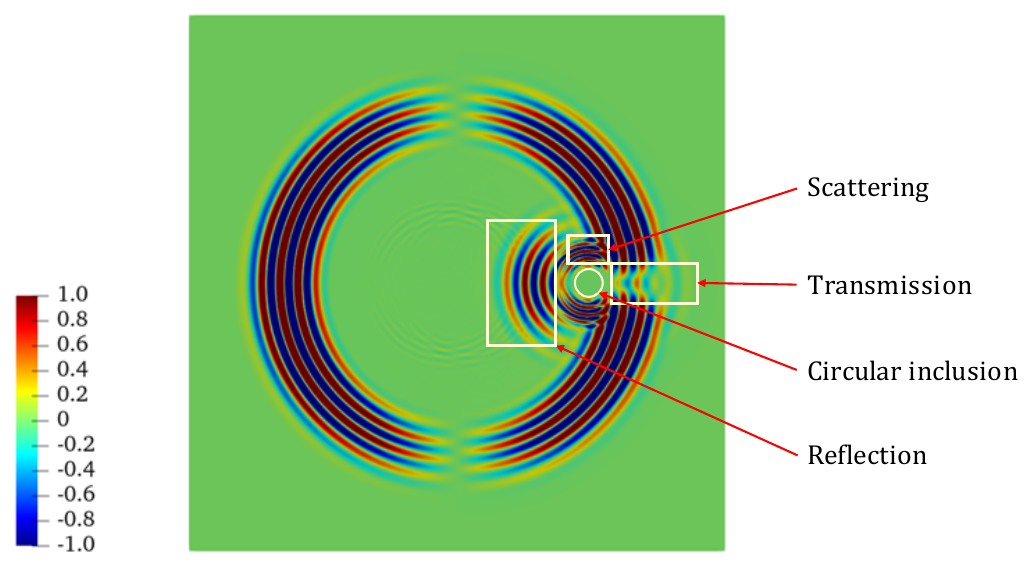}
		\caption{Illustration of the UGWs displacement field components $u_x$ in a solid plate}
		\label{fig:signal1}
	\end{subfigure}
	\begin{subfigure}[t]{0.48\textwidth}
		\centering
		\includegraphics[scale=0.8]{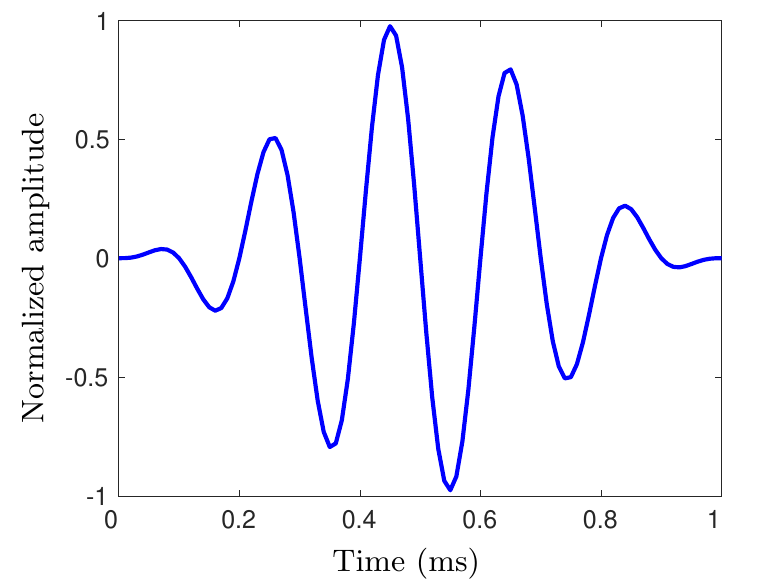}
		\caption{Five-cycle Hanning windowed Lamb wave burst signal with a center frequency of $5$ kHz}
		\label{fig:signal2}
	\end{subfigure}
	\begin{subfigure}[t]{0.48\textwidth}
		\centering
		\includegraphics[scale=0.8]{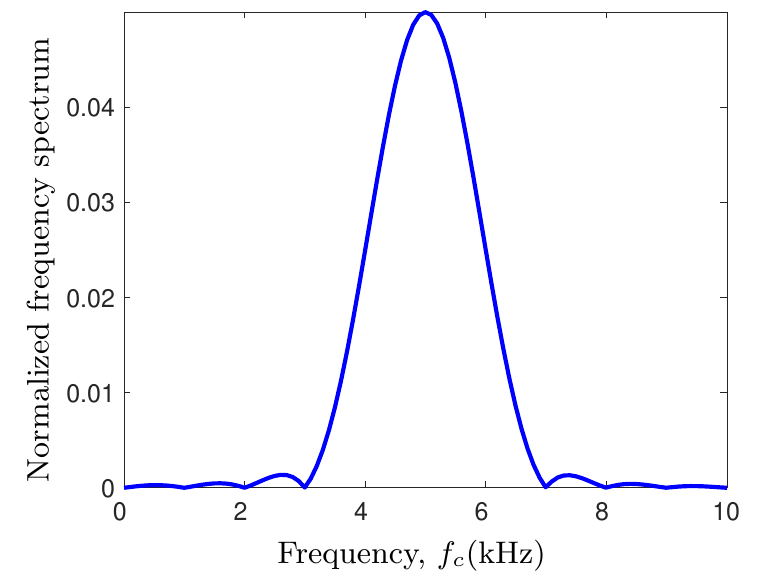}
		\caption{Frequency spectrum}
		\label{fig:signal3}
	\end{subfigure}
	\caption{The non-vanishing Lamb wave burst signal propagation in a solid plate. The displacement are normalized between $[-1,+1] $.}
	\label{fig:signal}
\end{figure}

The non-vanishing burst signal force is given by
\begin{eqnarray}
\begin{aligned}
f_{s}( x,t) =
\begin{bmatrix}
r(t)\cos(\varphi( x)) \\
r(t)\sin(\varphi( x)) 
\end{bmatrix}, \quad x \in \Omega_{f_{s}},
\end{aligned}
\label{eq6.1}
\end{eqnarray}
\vspace{0pt}
where $r(t)$ is radius of the non-vanishing signal at time $t$. The time evolution of $r(t)$ can be approximated by
\begin{eqnarray}
\begin{aligned}
r(t) := \left(\frac{1}{4}\cos\left(2\pi f_ct\right)\right)
\cdot
\left(\operatorname{sign}(t)-\operatorname{sign}\left(t-\frac{n_c}{f_c}\right)\right)
\cdot
\left(1-\cos\left(2\pi\frac{f_c}{n_c}t\right)\right),
\end{aligned}
\label{eq6.3}
\end{eqnarray}
where, carrier frequency $f_c=5$ kHz and modulation frequency $f_m=1$ kHz, which corresponds to $n_c=5$ cycles of the carrier signal in $[0,1] $ ms (see Figure \ref{fig:signal}). 
Futhermore, $ \Omega_{f_{s}} = \big\{ x_1, x_2 \in \Omega_s \hspace{1mm} \big\vert \hspace{2mm} x_1^2 + x_2^2 \le r_{f_{s}}^2\big\},$ and $\varphi( x)$ is given by
\begin{eqnarray}
\begin{aligned}
\varphi( x) = \begin{cases}
\displaystyle \arctan\bigg(\frac{ x_2}{ x_1}\bigg), & x_1 > 0 \wedge x_2 \ge 0,\\[1.25ex]
\displaystyle \frac{\pi}{2}, & x_1 = 0 \wedge x_2 > 0, \\[1.25ex]
\displaystyle \pi + \arctan\bigg(\frac{ x_2}{ x_1}\bigg), & x_1 < 0,\\[1.25ex]
\displaystyle \frac{3 \pi}{2}, & x_1 = 0 \wedge x_2 < 0,\\[1.25ex]
\displaystyle 2 \pi + \arctan\bigg(\frac{ x_2}{ x_1}\bigg), & x_1 > 0 \wedge x_2 < 0.
\end{cases}
\end{aligned}
\label{eq6.4}
\end{eqnarray}
%

\bibliographystyle{unsrt}
\bibliography{references}

\begin{thebibliography}{10}

\bibitem{SHMPWAS2014}
Victor Giurgiutiu.
\newblock {\em Structural Health Monitoring with Piezoelectric Wafer Active
  Sensors}.
\newblock Academic Press, 2014.

\bibitem{NDT2013}
Vistasp~M. Karbhari, editor.
\newblock {\em Non-Destructive Evaluation (NDE) of Polymer Matrix Composites}.
\newblock Woodhead Publishing, 2013.

\bibitem{RAMSHM2017}
He~Ren, Xi~Chen, and Yong Chen.
\newblock {\em Reliability Based Aircraft Maintenance Optimization and
  Applications}.
\newblock Academic Press, 2017.

\bibitem{CTSHM2011}
Srinivasan Gopalakrishnan, Massimo Ruzzene, and Sathyanarayana Hanagud.
\newblock {\em Computational Techniques for Structural Health Monitoring}.
\newblock Springer, 2011.

\bibitem{SHMAS2016}
Fuh-Gwo Yuan, editor.
\newblock {\em Structural Health Monitoring (SHM) in Aerospace Structures}.
\newblock Woodhead Publishing, Elsevier, 2016.

\bibitem{UGW2011}
Y.~Gharaibeh, R.~Sanderson, P.~Mudge, C.~Ennaceur, and W.. Balachandran.
\newblock Investigation of the behaviour of selected ultrasonic guided wave
  modes to inspect rails for long-range testing and monitoring.
\newblock {\em Proceedings of the Institution of Mechanical Engineers, Part F:
  Journal of Rail and Rapid Transit}, 225(3):311--324, 2011.

\bibitem{NTSHM2013}
Wieslaw Ostachowicz and J.~Alfredo G\"{u}emes, editors.
\newblock {\em New Trends in Structural Health Monitoring}.
\newblock Springer, 2013.

\bibitem{NSTSHM2011}
Subhas~Chandra Mukhopadhyay, editor.
\newblock {\em New Developments in Sensing Technology for Structural Health
  Monitoring}.
\newblock Springer, 2011.

\bibitem{NTVSHM2010}
Arnaud Deraemaeker and Keith Worden, editors.
\newblock {\em New Trends in Vibration Based Structural Health Monitoring}.
\newblock Springer-Wien-NewYork, 2010.

\bibitem{UGW2014}
Cara~A.C. Leckey, Matthew~D. Rogge, and F.~Raymond Parker.
\newblock Guided waves in anisotropic and quasi-isotropic aerospace composites:
  Three-dimensional simulation and experiment.
\newblock {\em Ultrasonics}, 54:385--394, 2014.

\bibitem{GW2017}
Matthieu Gresil, Adisorn Poohsai, and Neha Chandarana.
\newblock Guided wave propagation and damage detection in composite pipes using
  piezoelectric sensors.
\newblock {\em Procedia Engineering}, 188:148--155, 2017.

\bibitem{UGW2009}
Fei Yan, Roger~L. Royer~jr, and Joseph~L. Rose.
\newblock Ultrasonic guided wave imaging techniques in structural health
  monitoring.
\newblock {\em Journal of Intelligent Material Systems and Structures},
  21(3):377--384, 2009.

\bibitem{UGW2010}
Piervincenzo Rizzo, Jian-Gang Han, and Xiang-Lei Nit.
\newblock Structural health monitoring of immersed structures by means of
  guided ultrasonic waves.
\newblock {\em Journal of Intelligent Material Systems and Structures},
  21(14):1397--1407, 2010.

\bibitem{LW2016}
A~De Luca, Z.~Sharif-Khodaeib, M.~H. Aliabadib, and F.~Caputo.
\newblock Numerical simulation of the lamb wave propagation in impacted cfrp
  laminate.
\newblock {\em Procedia Engineering}, 167:109--115, 2016.

\bibitem{LAMB1917}
Horace Lamb.
\newblock On waves in an elastic plate.
\newblock {\em Proceedings of the Royal Society A}, 93:114--128, 1917.

\bibitem{WAVE2014}
Joseph~l. Rose.
\newblock {\em Ultrasonic Guided Waves in Solid Media}.
\newblock Cambridge University Press, 2014.

\bibitem{LEE2014}
Jung-Ryul Lee, Jae-Kyeong Jang, and Cheol-Won Kong.
\newblock Fully noncontact wave propagation imaging in an immersed metallic
  plate with a crack.
\newblock {\em Shock and Vibration}, 5:895693/1--8, 2014.

\bibitem{REVIEW2007}
Ajay Raghavan and Carlos E.~S. Cesnik.
\newblock Review of guided-wave structural health monitoring.
\newblock {\em The Shock and Vibration Digest}, 39(2):91--114, 2007.

\bibitem{REVIEW2010}
Guoliang Huang, Fei Song, and Xiaodong Wang.
\newblock Quantitative modeling of coupled piezo-elastodynamic behavior of
  piezoelectric actuators bonded to an elastic medium for structural health
  monitoring: A review.
\newblock {\em Sensors}, 10:3681--3702, 2010.

\bibitem{REVIEW2016}
Mira Mitra and S.~Gopalakrishnan.
\newblock Guided wave based structural health monitoring: A review.
\newblock {\em Smart Materials and Structures}, 25(5):053001/1--27, 2016.

\bibitem{LWSHM2018}
Rolf Lammering, Ulrich Gabbert, Michael Sinapius, Thomas Schuster, and Peter
  Wierach, editors.
\newblock {\em Lamb-Wave Based Structural Health Monitoring in Polymer
  Composites}.
\newblock Springer, 2018.

\bibitem{EbnaHaiDE2017}
Bhuiyan Shameem~Mahmood Ebna~Hai.
\newblock {\em Finite element approximation of ultrasonic wave propagation
  under fluid-structure interaction for structural health monitoring systems}.
\newblock PhD thesis, Faculty of Mechanical Engineering, Helmut Schmidt
  University-University of the Federal Armed Forces Hamburg, Germany, 2017.

\bibitem{EbnaHaiTS2019}
Bhuiyan Shameem~Mahmood Ebna~Hai and Markus Bause.
\newblock Numerical study and comparison of alternative time discretization
  schemes for an ultrasonic guided wave propagation problem coupled with
  fluid--structure interaction.
\newblock {\em Computers \& Mathematics with Applications}, 78(9):2867--2885,
  2019.

\bibitem{EbnaHaiWpFSI2019}
Bhuiyan Shameem~Mahmood Ebna~Hai, Markus Bause, and Paul Kuberry.
\newblock Modeling and simulation of ultrasonic guided waves propagation in the
  fluid-structure domain by a monolithic approach.
\newblock {\em Journal of Fluids and Structures}, 88:100--121, 2019.

\bibitem{EbnaHaieXFSI2019}
Bhuiyan Shameem~Mahmood Ebna~Hai, Markus Bause, and Paul~Allen Kuberry.
\newblock Modeling concept and numerical simulation of ultrasonic wave
  propagation in a moving fluid-structure domain based on a monolithic
  approach.
\newblock {\em Applied Mathematical Modelling}, 75:916--939, 2019.

\bibitem{TEMP2008}
Francesco Lanza~di Scalea and Salvatore Salamone.
\newblock Temperature effects in ultrasonic lamb wave structural health
  monitoring systems.
\newblock {\em The Journal of the Acoustical Society of America}, 124(1), 2008.

\bibitem{TEMP2009}
Salvatore Salamone, Ivan Bartoli, Francesco Lanza~Di Scalea, and Stefano
  Coccia.
\newblock Guided-wave health monitoring of aircraft composite panels under
  changing temperature.
\newblock {\em Journal of Intelligent Material Systems and Structures},
  20(9):1079--1090, 2009.

\bibitem{williams2006gaussian}
Carl~Edward Rasmussen and Christopher~KI Williams.
\newblock {\em Gaussian processes for machine learning}.
\newblock MIT press Cambridge, MA, 2006.

\bibitem{krige1951statistical}
Daniel~G Krige.
\newblock A statistical approach to some basic mine valuation problems on the
  {W}itwatersrand.
\newblock {\em Journal of the Southern African Institute of Mining and
  Metallurgy}, 52(6):119--139, 1951.

\bibitem{koch2015efficient}
Patrick Koch, Tobias Wagner, Michael~TM Emmerich, Thomas B{\"a}ck, and Wolfgang
  Konen.
\newblock Efficient multi-criteria optimization on noisy machine learning
  problems.
\newblock {\em Applied Soft Computing}, 29:357--370, 2015.

\bibitem{emmerich2006single}
Michael~TM Emmerich, Kyriakos~C Giannakoglou, and Boris Naujoks.
\newblock Single-and multiobjective evolutionary optimization assisted by
  {G}aussian random field metamodels.
\newblock {\em IEEE Transactions on Evolutionary Computation}, 10(4):421--439,
  2006.

\bibitem{wang2007review}
G~Gary Wang and Songqing Shan.
\newblock Review of metamodeling techniques in support of engineering design
  optimization.
\newblock {\em Journal of Mechanical design}, 129:370--380, 2007.

\bibitem{forrester2008engineering}
Alexander Forrester, Andras Sobester, and Andy Keane.
\newblock {\em Engineering design via surrogate modelling: a practical guide}.
\newblock John Wiley \& Sons, 2008.

\bibitem{goodfellow2016deep}
Ian Goodfellow, Yoshua Bengio, and Aaron Courville.
\newblock {\em Deep learning}.
\newblock MIT press, 2016.

\bibitem{baker2019workshop}
Nathan Baker, Frank Alexander, Timo Bremer, Aric Hagberg, Yannis Kevrekidis,
  Habib Najm, Manish Parashar, Abani Patra, James Sethian, Stefan Wild, Karen
  Willcox, and Steven Lee.
\newblock Workshop report on basic research needs for scientific machine
  learning: Core technologies for artificial intelligence.
\newblock Technical report, USDOE Office of Science (SC), Washington, DC
  (United States), 2019.

\bibitem{raissi2017physics}
Maziar Raissi, Paris Perdikaris, and George~Em Karniadakis.
\newblock Physics informed deep learning (part {I}): {Data-driven} solutions of
  nonlinear partial differential equations.
\newblock {\em arXiv preprint arXiv:1711.10561}, 2017.

\bibitem{raissi2018hidden}
Maziar Raissi and George~Em Karniadakis.
\newblock Hidden physics models: Machine learning of nonlinear partial
  differential equations.
\newblock {\em Journal of Computational Physics}, 357:125--141, 2018.

\bibitem{sun2021physics}
Jian Sun, Kristopher~A Innanen, and Chao Huang.
\newblock Physics-guided deep learning for seismic inversion with hybrid
  training and uncertainty analysis.
\newblock {\em Geophysics}, 86(3):1--64, 2021.

\bibitem{lino2020simulating}
Mario Lino, Chris Cantwell, Stathi Fotiadis, Eduardo Pignatelli, and Anil
  Bharath.
\newblock Simulating surface wave dynamics with convolutional networks.
\newblock {\em arXiv preprint arXiv:2012.00718}, 2020.

\bibitem{sorteberg2018approximating}
Wilhelm~E Sorteberg, Stef Garasto, Alison~S Pouplin, Chris~D Cantwell, and
  Anil~A Bharath.
\newblock Approximating the solution to wave propagation using deep neural
  networks.
\newblock {\em arXiv preprint arXiv:1812.01609}, 2018.

\bibitem{moseley2018fast}
Benjamin Moseley, Andrew Markham, and Tarje Nissen-Meyer.
\newblock Fast approximate simulation of seismic waves with deep learning.
\newblock {\em arXiv preprint arXiv:1807.06873}, 2018.

\bibitem{yang2019deep}
Wenming Yang, Xuechen Zhang, Yapeng Tian, Wei Wang, Jing-Hao Xue, and Qingmin
  Liao.
\newblock Deep learning for single image super-resolution: A brief review.
\newblock {\em IEEE Transactions on Multimedia}, 21(12):3106--3121, 2019.

\bibitem{dong2014learning}
Chao Dong, Chen~Change Loy, Kaiming He, and Xiaoou Tang.
\newblock Learning a deep convolutional network for image super-resolution.
\newblock In {\em European Conference on Computer Vision}, pages 184--199.
  Springer, 2014.

\bibitem{dong2015image}
Chao Dong, Chen~Change Loy, Kaiming He, and Xiaoou Tang.
\newblock Image super-resolution using deep convolutional networks.
\newblock {\em IEEE Transactions on Pattern Analysis and Machine Intelligence},
  38(2):295--307, 2015.

\bibitem{stengel2019physics}
Karen Stengel, Andrew Glaws, and Ryan King.
\newblock Physics-informed super resolution of climatological wind and solar
  resource data.
\newblock {\em AGUFM}, 2019:A43E--04, 2019.

\bibitem{zhu2017wave}
Weiqiang Zhu, Yixiao Sheng, and Yi~Sun.
\newblock Wave-dynamics simulation using deep neural networks.

\bibitem{melville2018structural}
Joseph Melville, K~Supreet Alguri, Chris Deemer, and Joel~B Harley.
\newblock Structural damage detection using deep learning of ultrasonic guided
  waves.
\newblock In {\em AIP Conference Proceedings}, volume 1949, page 230004. AIP
  Publishing LLC, 2018.

\bibitem{khurjekar2019accounting}
Ishan~D Khurjekar and Joel~B Harley.
\newblock Accounting for physics uncertainty in ultrasonic wave propagation
  using deep learning.
\newblock {\em arXiv preprint arXiv:1911.02743}, 2019.

\bibitem{rautela2021ultrasonic}
Mahindra Rautela and S~Gopalakrishnan.
\newblock Ultrasonic guided wave based structural damage detection and
  localization using model assisted convolutional and recurrent neural
  networks.
\newblock {\em Expert Systems with Applications}, 167:114189, 2021.

\bibitem{willard2020integrating}
Jared Willard, Xiaowei Jia, Shaoming Xu, Michael Steinbach, and Vipin Kumar.
\newblock Integrating physics-based modeling with machine learning: A survey.
\newblock {\em arXiv preprint arXiv:2003.04919}, 2020.

\bibitem{zhang2019quantifying}
Dongkun Zhang, Lu~Lu, Ling Guo, and George~Em Karniadakis.
\newblock Quantifying total uncertainty in physics-informed neural networks for
  solving forward and inverse stochastic problems.
\newblock {\em Journal of Computational Physics}, 397:108850, 2019.

\bibitem{hinton2012improving}
Geoffrey~E Hinton, Nitish Srivastava, Alex Krizhevsky, Ilya Sutskever, and
  Ruslan~R Salakhutdinov.
\newblock Improving neural networks by preventing co-adaptation of feature
  detectors.
\newblock {\em arXiv preprint arXiv:1207.0580}, 2012.

\bibitem{de2020transfer}
Subhayan De, Jolene Britton, Matthew Reynolds, Ryan Skinner, Kenneth Jansen,
  and Alireza Doostan.
\newblock On transfer learning of neural networks using bi-fidelity data for
  uncertainty propagation.
\newblock {\em International Journal for Uncertainty Quantification},
  10(6):543--573, 2020.

\bibitem{de2020uncertainty}
Subhayan De.
\newblock Uncertainty quantification of locally nonlinear dynamical systems
  using neural networks.
\newblock {\em arXiv preprint arXiv:2008.04598}, 2020.

\bibitem{Miyanawala-FSI-ML2018}
T.~P. Miyanawala and R.~Jaiman.
\newblock A hybrid data-driven deep learning technique for fluid-structure
  interaction.
\newblock {\em arXiv: Computational Physics}, 2018.

\bibitem{Whisenant-FSI-ML2020}
Matthew~J. Whisenant and Kivanc Ekici.
\newblock Galerkin-free technique for the reduced-order modeling of
  fluid-structure interaction via machine learning.
\newblock In {\em AIAA Scitech 2020 Forum}. AIAA, 2020.

\bibitem{WickPFF2016}
Thomas Wick.
\newblock Coupling fluid--structure interaction with phase-field fracture.
\newblock {\em Journal of Computational Physics}, 327:67--96, 2016.

\bibitem{Richter2017book}
Thomas Richter.
\newblock {\em Fluid-structure Interactions: Models, Analysis and Finite
  Elements.}, volume 118.
\newblock Springer, 2017.

\bibitem{paszke2017automatic}
Paszke Adam, Gross Sam, Chintala Soumith, Chanan Gregory, Yang Edward,
  D~Zachary, Lin Zeming, Desmaison Alban, Antiga Luca, and Lerer Adam.
\newblock Automatic differentiation in {PyTorch}.
\newblock In {\em Proceedings of Neural Information Processing Systems}, 2017.

\bibitem{tai2017image}
Ying Tai, Jian Yang, and Xiaoming Liu.
\newblock Image super-resolution via deep recursive residual network.
\newblock In {\em Proceedings of the IEEE Conference on Computer Vision and
  Pattern Recognition}, pages 3147--3155, 2017.

\bibitem{bottou2018optimization}
L{\'e}on Bottou, Frank~E Curtis, and Jorge Nocedal.
\newblock Optimization methods for large-scale machine learning.
\newblock {\em SIAM Review}, 60(2):223--311, 2018.

\bibitem{kingma2014adam}
Diederik~P Kingma and Jimmy Ba.
\newblock Adam: A method for stochastic optimization.
\newblock {\em arXiv preprint arXiv:1412.6980}, 2014.

\bibitem{de2019topology}
Subhayan De, Jerrad Hampton, Kurt Maute, and Alireza Doostan.
\newblock Topology optimization under uncertainty using a stochastic
  gradient-based approach.
\newblock {\em Structural and Multidisciplinary Optimization},
  62(5):2255--2278, 2020.

\bibitem{wang2004image}
Zhou Wang, Alan~C Bovik, Hamid~R Sheikh, and Eero~P Simoncelli.
\newblock Image quality assessment: from error visibility to structural
  similarity.
\newblock {\em IEEE Transactions on Image Processing}, 13(4):600--612, 2004.

\bibitem{ahn2018fast}
Namhyuk Ahn, Byungkon Kang, and Kyung-Ah Sohn.
\newblock Fast, accurate, and lightweight super-resolution with cascading
  residual network.
\newblock In {\em Proceedings of the European Conference on Computer Vision
  (ECCV)}, pages 252--268, 2018.

\bibitem{Agustsson_2017_CVPR_Workshops}
Eirikur Agustsson and Radu Timofte.
\newblock {NTIRE} 2017 challenge on single image super-resolution: Dataset and
  study.
\newblock In {\em The IEEE Conference on Computer Vision and Pattern
  Recognition (CVPR) Workshops}, 2017.

\bibitem{WickMMT2011}
Thomas Wick.
\newblock Fluid-structure interactions using different mesh motion techniques.
\newblock {\em Computers \& Structures}, 89(13-14):1456--1467, 2017.

\bibitem{UMFPACK}
Timothy~A Davis and Iain~S Duff.
\newblock An unsymmetric-pattern multifrontal method for sparse lu
  factorization.
\newblock {\em SIAM Journal on Matrix Analysis and Applications},
  18(1):140--158, 1997.

\bibitem{EbnaHaieXFSI2017}
Bhuiyan Shameem~Mahmood Ebna~Hai and Markus Bause.
\newblock Finite element approximation of fluid-structure interaction with
  coupled wave propagation.
\newblock {\em PAMM - Proceedings in Applied Mathematics and Mechanics},
  17(1):511--512, 2017.

\bibitem{DOpElib}
Christian Goll, Thomas Wick, and Winnifried Wollner.
\newblock Dopelib: Differential equations and optimization environment; a goal
  oriented software library for solving pdes and optimization problems with
  pdes.
\newblock {\em Archive of Numerical Software}, 5(2), 2017.

\bibitem{deal}
W.~Bangerth, R.~Hartmann, and G.~Kanschat.
\newblock \texttt{deal.II} -- a general purpose object-oriented finite element
  library.
\newblock {\em ACM Transactions on Mathematical Software}, 33(4):24/1--27,
  2007.

\bibitem{de2020bi}
Subhayan De, Kurt Maute, and Alireza Doostan.
\newblock Bi-fidelity stochastic gradient descent for structural optimization
  under uncertainty.
\newblock {\em Computational Mechanics}, 66(4):745--771, 2020.

\bibitem{Ciarlet1978book}
P.~G. Ciarlet.
\newblock {\em The finite element method for elliptic problems.}, volume~4.
\newblock North-Holland, Amsterdam, 1978.

\bibitem{Ciarlet1974BC}
P.~G. Ciarlet and P.~A. Raviart.
\newblock A mixed finite element method for the biharmonic equation.
\newblock {\em In: C. de Boor (ed.), Mathematical aspects of finite elements in
  partial differential equations.}, pages 125--145, 1974.

\end{thebibliography}

\end{document}